\documentclass[a4paper]{article}

\usepackage{RR}
\usepackage{hyperref} 
\usepackage{graphicx} 
\usepackage[dvips]{epsfig} 

\RRdate{February 2007}
 
\newtheorem{lemma}{Lemma} 
\newtheorem{defin}{Definition} 
\newtheorem{conse}{Corollary} 
\newtheorem{theor}{Theorem} 
\newtheorem{remar}{Remark} 
 
\newtheorem{proposition}{Proposition} 
\newtheorem{example}{Example} 
 
\newcommand{\defi}{\stackrel{\triangle}{=}} 
\def\R{\mathop{\rm I\kern -0.20em R}\nolimits} 
\def\blacksquare{\hbox{\vrule width 10pt height 10pt depth 0pt}}

\RRauthor{
Konstantin Avrachenkov\thanks{INRIA Sophia Antipolis, France, e-mail: 
k.avrachenkov@sophia.inria.fr .}
\and 
Urtzi Ayesta\thanks{LAAS-CNRS, France, e-mail: urtzi@laas.fr .}
\and
Alexei Piunovskiy\thanks{ University of Liverpool, UK, e-mail: 
piunov@liverpool.ac.uk}
}

\authorhead{Avrachenkov, Ayesta \& Piunovskiy}

\RRtitle{Convergence et choix optimal de la taille du buffer du routeur
pour le contr\^ole de congestion AIMD}
\RRetitle{Convergence and Optimal Buffer Sizing for Window Based 
AIMD Congestion Control}
\titlehead{Convergence and Optimal Buffer Sizing for 
AIMD Congestion Control}
\RRnote{The work was supported by France Telecom R\&D Grant 
``Mod\'elisation et Gestion du Trafic R\'eseaux Internet''
no. 42937433.}

\RRresume{Nous \'etudions l'interaction entre le contr\^ole de congestion
AIMD (Additive Increase Multiplicative Decrease) et le routeur avec le buffer 
de type Drop Tail. Nous consid\'erons ce probl\`eme dans le cadre des mod\`eles 
hybrides d\'eterministes. D'abord, nous prouvons que le mod\`ele hybride 
de l'interaction entre la c\^ontrole de congestion AIMD et le routeur de 
goulot d'\'etranglement converge toujours \`a un comportement cyclique. 
Nous caract\'erisons les cycles. Des conditions n\'ecessaires et suffisantes 
pour l'absence des sauts multiples de la fen\^etre de congestion dans 
le m\^eme cycle sont obtenues. Puis, nous proposons un cadre analytique 
pour le choix optimal de la taille du buffer du routeur. Nous formulons 
le probl\`eme du choix optimal de la taille du buffer du routeur
comme probl\`eme d'optimisation multi-crit\`ere, dans lequel la fonction 
de Lagrange correspond \`a une combinaison lin\'eaire du taux moyen de transmission 
et le d\'elai moyen dans le buffer. La solution au probl\`eme d'optimisation 
fournit davantage d'\'evidence que la taille du buffer du routeur doit \^etre 
r\'eduite en pr\'esence de l'agr\'egation du trafic. Nos r\'esultats analytiques 
sont confirm\'es par des simulations effectu\'ees avec Simulink et le simulateur NS.}

\RRabstract{We study the interaction between the AIMD (Additive 
Increase Multiplicative Decrease) congestion control and a 
bottleneck router with Drop Tail buffer. We consider the problem in 
the framework of deterministic hybrid models. First, we show that 
the hybrid model of the interaction between the AIMD congestion 
control and bottleneck router always converges to a cyclic behavior. 
We characterize the cycles. Necessary and sufficient conditions for 
the absence of multiple jumps of congestion window in the same cycle 
are obtained. Then, 
we propose an analytical framework for the optimal choice of the 
router buffer size. We formulate the problem of the optimal router 
buffer size as a multi-criteria optimization problem, in which the 
Lagrange function corresponds to a linear combination of the average 
goodput and the average delay in the queue. The solution to the 
optimization problem provides further evidence that the buffer size 
should be reduced in the presence of traffic aggregation. Our 
analytical results are confirmed by simulations performed with 
Simulink and the NS simulator.}

\RRmotcle{le mod\`ele de TCP/IP, choix optimal de la taille du buffer,
mod\`eles hybrides d\'eterministes, l'optimisation multi-crit\`ere}

\RRkeyword{TCP/IP modeling, router buffer sizing, deterministic hybrid model 
approach, multi-criteria optimization}

\RRprojet{Maestro} 

\RRtheme{\THCom} 

\URSophia

\begin{document} 
 
\makeRR

\section{Introduction} 
 
Most traffic in the Internet is governed by TCP/IP (Transmission 
Control Protocol and Internet Protocol) \cite{rfc2581,J88}. Data 
packets of an Internet connection travel from a source node to a 
destination node via a series of routers. Some routers, particularly 
edge routers, experience periods of congestion when packets spend a 
non-negligible time waiting in the router buffers to be transmitted 
over the next hop. TCP protocol tries to adjust the sending rate of 
a source to match the available bandwidth along the path. During the 
principle Congestion Avoidance phase the current TCP New Reno 
version uses AIMD (Additive Increase Multiplicative Decrease) binary 
feedback congestion control scheme. In the absence of congestion 
signals from the network TCP increases congestion window linearly in 
time, and upon the reception of a congestion signal TCP reduces the 
congestion window by a multiplicative factor. Congestion signals can 
be either packet losses or ECN (Explicit Congestion Notifications) 
\cite{rfc3168}. At the present state of the Internet, nearly all 
congestion signals are generated by packet losses. Packets can be 
dropped either when the router buffer is full or when AQM (Active 
Queue Management) scheme is employed \cite{FJ93}. Given an ambiguity 
in the choice of the AQM parameters \cite{CJOD01,MBDL99}, so far AQM 
is rarely used in practice. On the other hand, in the basic Drop 
Tail routers, the buffer size is the only one parameter to tune 
apart of the router capacity. In fact, the buffer size is one of few 
parameters of the TCP/IP network that can be managed by network 
operators. This makes the choice of the router buffer size a very 
important problem in the TCP/IP network design. 
 
The paper is composed of two principle parts. In the first part 
(Sections~\ref{sec:model}-\ref{sec:multijumps}) we analyze the 
interaction between the AIMD congestion control and the bottleneck 
router with Drop Tail buffer. This interaction can be adequately 
described by hybrid modeling approach. There are several hybrid 
models of the interaction between TCP and the bottleneck router 
\cite{AAP05,BHLO03,HBOL01}. Here we analyze the model of 
\cite{HBOL01}. To our opinion, this model takes into account all 
essential details of TCP and at the same time leads to a tractable 
analysis. We show that the system always converges to a limiting 
behavior. In particular, we demonstrate that two different limiting 
regimes can coexist and the convergence to one or to the other 
depends on the initial conditions. Then, we provide necessary and 
sufficient conditions for the absence of subsequent packet losses. 
The absence of subsequent packet losses benefits the TCP performance 
as well as the quality of service for end users. We note that in 
\cite{HBOL01} there is no characterization of limiting regimes. 
Furthermore, in \cite{HBOL01} only a sufficient condition for the 
absence of multiple jumps was obtained and the sufficient condition 
of \cite{HBOL01} is loose for some values of the decrease factor. 
 
In the second part of the paper 
(Sections~\ref{sec:pareto}-\ref{sec:bmin}) we study the optimal 
choice of the buffer size in the bottleneck routers. There are some 
empirical rules for the choice of the router buffer size. The first 
proposed rule of thumb for the choice of the router buffer size was 
to choose the buffer size equal to the BDP (Bandwidth-Delay Product) 
of the outgoing link \cite{VS94}. This recommendation is based on 
very approximative considerations and it can be justified only when 
a router is saturated with a single long-lived TCP connection. The 
next apparent question to ask was how one should set the buffer size 
in the case of several competing TCP connections. In \cite{AAANB02} 
it was observed that the utilization of a link improves very fast 
with the increase of the buffer size until a certain threshold 
value. After that threshold value the further increase of the buffer 
size does not improve the link utilization but increases the 
queueing delay. Then, two contradictory guidelines for the choice of 
the buffer size have been proposed. In \cite{M00} a 
connection-proportional buffer size allocation is proposed, whereas 
in \cite{AKM04} it was suggested that the buffer size should be set 
to the BDP of the outgoing link divided by the square root of the 
number of TCP connections. A rationale for the former recommendation 
is that in order to avoid a high loss rate the buffer must 
accommodate at least few packets from each connection. And a 
rationale for the latter recommendation is based on the reduction of 
the synchronization of TCP connections when the number of 
connections increases. Then, \cite{AKM04,M00} were followed by two 
works \cite{DJD05,GKT04} which try to reconcile these two 
contradictory approaches. In particular, the authors of \cite{DJD05} 
recommend to follow the rule of \cite{AKM04} for a relatively small 
number of long-lived connections and, when the number of long-lived 
bottlenecked connections is large, to switch to the 
connection-proportional allocation. One of the main conclusions of 
\cite{GKT04} is that there are no clear criteria for the 
optimization of the buffer size. Then, the author of \cite{GKT04} 
proposed a general avenue for research on the router buffer sizing: 
``Find the link buffer size that accommodates both TCP and UDP 
traffic.'' We note that UDP (User Datagram Protocol) \cite{rfc768} 
does not use any congestion control and reliable retransmission and 
it is mostly employed for delay sensitive applications such as 
Internet Telephony. We refer the interested reader to \cite{Vu07} 
and references therein for more information on the problem of 
optimal choice of buffer size. 
 
All the above mentioned works on the router buffer sizing are based 
on quite rough approximations and strictly speaking do not take into account 
the feedback nature of TCP protocol. Here we propose a mathematically solid framework 
to analyze the interaction of TCP with the finite buffer of an IP router. In particular, 
we state a criterion for the choice of the optimal buffer size in a mathematical form. 
Our optimization criterion can be considered as a mathematical formalization of 
the lingual criterion proposed in \cite{GKT04}. Furthermore, the Pareto set obtained 
for our model allows us to dimension the IP router buffer size to accommodate both 
data traffic and real time traffic. 
 
All proofs are provided in the Appendix.

\section{Mathematical model} 
\label{sec:model} 
 
The window based binary feedback congestion control can be described by 
two functions $f(w)$ and $G(w)$. Function $f(w)$ defines the increase profile 
of the congestion window and function $G(w)$ represents the reduction of the 
congestion window upon the reception of congestion notification. Namely, in the 
absence of congestion notification the evolution of congestion window $w(t)$ 
is described by the differential equation 
\begin{equation} 
\label{wind} \frac{dw}{dt}=\frac{f(w)}{T+x(t)/\mu}, 
\end{equation} 
where $T$ is the two way propagation delay, $x(t)$ is the amount of 
data in the bottleneck queue and $\mu$ is the capacity of the 
bottleneck router. We note that $T+x(t)/\mu$ corresponds to the 
Round Trip Time (RTT) when the amount of the enqueued data in the 
bottleneck router buffer is $x(t)$ at time moment $t$. Thus, 
function $f(w)$ determines the increase of the congestion window per 
one Round Trip Time. The sending rate $\lambda(t)$ of the window 
based congestion control is given by 
\begin{equation} 
\label{lambda} \lambda(t)=\frac{w(t)}{T+x(t)/\mu}. 
\end{equation} 
We would like to emphasize that here the time parameter $t$ corresponds to 
the local time observed at the router. 
 
We study a Drop Tail buffer with size $B$. If $x(t)<B$, the congestion 
window $w$ increases according to (\ref{wind}). When $x$ reaches $B$ 
at time $t^*$, i.e. $x(t^*)=B$, the buffer starts to overflow. The 
overflow of the buffer will be noticed by the sender only after the 
time delay $\delta=T+B/\mu$. Upon the reception of the congestion 
signal at time $t^*+\delta$, the congestion window is reduced 
according to 
\begin{equation} 
\label{Gfunc} 
w(t^*+\delta+0)=G(w(t^*+\delta-0)). 
\end{equation} 
As we shall see below, $w$ (resp., $\lambda$) can represent either a congestion window 
(resp., sending rate) for a single TCP connection or a total window (resp., total rate) 
of several TCP connections. 
 
Consider $n$ long-lived AIMD 
TCP connections that share a bottleneck router. Denote by $w_i(t)$ 
the instantaneous congestion window of connection $i=1,...,n$ at time $t\in[0,\infty)$. 
In the case of the AIMD congestion control, if $x<B$ the evolution of 
the congestion window $w_i$ is given by differential equation (\ref{wind}) 
with $f(w)=m_i=const$. 
If we restrict ourselves to the symmetric case $T_i=T=const$ and $m_i=m_0=const$, 
the sum of all congestion windows $w(t)=\sum_{i=1}^nw_i(t)$ also satisfies 
differential equation (\ref{wind}) with $f(w)=m$, where $m=n m_0$. 
Namely,  
we have 
\begin{equation} 
\label{wind_AIMD} \frac{dw}{dt}=\frac{m}{T+x(t)/\mu}, 
\end{equation} 
\begin{equation} 
\label{x_evol_AIMD} 
\frac{dx}{dt} = 
\left\{ \begin{array}{ll} 
\lambda(t)-\mu, & \mbox{if} \ 0<x(t)<B, \ \mbox{or }  x(t)=0 \ \mbox{and} \ \lambda(t) \ge \mu, \ \mbox{or } 
\ x(t)=B \ \mbox{and} \ \lambda(t) \le \mu;\\ 
0 & \mbox{otherwise,} 
\end{array} \right. 
\end{equation} 
where $\lambda(t)$ is given by (\ref{lambda}). 
And if $x(t^*)=B$ at some time moment $t^*$, the congestion window is decreased 
multiplicatively after the information propagation delay $\delta=T+B/\mu$ as follows: 
\begin{equation} 
\label{Gfunc_AIMD} 
w(t^*+\delta+0)=\beta^k w(t^*+\delta-0), 
\end{equation} 
and consequently, $G(w)=\beta^k w$ for the AIMD case. Usually, $k=1$, but 
sometimes it is necessary to send several congestion signals in order to 
reduce the sending rate below the transmission capacity of the bottleneck 
router. 
 
Since we consider the case of equal propagation delays, the 
synchronization phenomenon takes place \cite{FJ93}, and 
consequently, the total sending rate is also reduced by the factor 
$\beta^k$. For instance, in TCP New Reno version the reduction 
factor $\beta$ is equal to one half. 
 
Let us make the change of time scale according to 
$$ 
ds \defi \frac{dt}{T+x(t)/\mu}. 
$$ 
and the change of variables: 
$$ 
v \defi w/m, \quad y \defi x/m. 
$$ 
The new time $s$ can be viewed as a counter for Round Trip Times. 
Now the dynamics of the system between the jumps is described 
by equations 
\begin{equation} 
\label{dyn_v} 
\frac{dv}{ds}=1, 
\end{equation} 
\begin{equation} 
\label{dyn_y} 
\frac{dy}{ds}=\left\{ 
\begin{array}{ll} 
v(t)-y(t)-q, & \mbox{if} \ 0<y(t)<b, \ \mbox{or } y(t)=0 \ \mbox{and} \ v(t)\ge q, \ \mbox{or } 
 y(t)=b \ \mbox{and} \ v(t)\le q+b;\\ 
0 & \mbox{otherwise,} 
\end{array} \right. 
\end{equation} 
where $q=\mu T/m$ is the maximal number of packets that can be fit 
in the pipe, in other words Bandwidth-Delay Product (BDP) in 
packets, and $b=B/m$ is the maximal number of packets that can be 
fit in the router buffer. Let $s^*$ be the moment in the new time scale 
when component $y$ reaches value $b$. Then, equation 
(\ref{Gfunc_AIMD}) is transformed to 
\begin{equation} 
\label{vreduction} 
v(s^*+1+0) = \beta^k v(s^*+1-0), 
\end{equation} 
where $k=\min \{ i: \beta^i v(s^*+1-0)<b+q\}$. 
 
\begin{remar} 
Because of the delay in the information propagation, the congestion 
window is reduced after the delay $\delta=T+B/\mu$ in the original 
time scale, or, equivalently, after $1$ time unit in the new time 
scale $s$. The value of $k$ is such that, after sending $k$ 
congestion signals, the amount of data $x$ $($and $y$ $)$ starts to 
decrease. 
\end{remar} 
 
\section{Convergence of the system trajectories} 
\label{sec:converge} 
 
The dynamics is defined by three parameters $\beta, q$, and $b$, and 
the system trajectory remains in the region $\Omega=\{0\le y\le 
b,~v>0\}$, provided the initial condition is there. 
 
Suppose a trajectory starts at $s=0$ from initial condition $y_0=b$, 
$\beta(q+b)\le v_0<b+q$,\footnote{Initial conditions outside the region 
$[\beta(q+b),q+b)$ are of no interest because, after the very first 
(multiple) jump we have $v(s^*+1+0)\in[\beta(q+b),q+b)$.} 
and $s^*$ is the first moment when $y(s^*)=b$. 
Let $v_1=v(s^*+1+0).$ We introduce mapping $\varphi$ 
such that $v_1\defi\varphi(v_0)$. Consider 
the iterations $v_{i+1}\defi\varphi(v_i)$, $i=0,1,...$. 
 
\begin{theor} 
\label{prop_stab} There exists 
$\lim_{i\to\infty}v_i=V(v_0)$ with 
\begin{equation} 
\label{ee9} 
V(v)=\left\{\begin{array}{ll} 
V_1, & \mbox{ if }v\in[\beta(q+b),d]; \\ 
V_2, & \mbox{ if }v\in(d,q+b), 
\end{array}\right. 
\end{equation} 
for some constant $d$. 
In particular, one of the above intervals can be empty. 
\end{theor} 
 
\begin{defin} 
\label{def_cycle} Suppose the trajectory starting at $s=0$ from 
initial condition $y_0=b$, $v_0<b+q$ reaches the same point, for the 
first time, at some time moment $S\ge 1$. Then this finite 
trajectory is called a {\em cycle}. A cycle with component $y$ 
remaining zero for a positive time interval is called {\em clipped} 
$($see Figure~\ref{fig:clipped-2-cycle}$)$. If a cycle touches the axis $y=0$ only 
at a single point, we call such cycle {\em critical} 
$($see Figure~\ref{fig:critical-2-cycle}$)$. 
\end{defin} 
 
\begin{conse}[from Theorem~\ref{prop_stab}] 
\label{cor_simple_cycle} Any cycle has a single time moment, when a 
(multiple) jump occurs. 
\end{conse} 
 
The number $k$ of instant jumps of component $v$ is called {\sl a 
cycle order}. We call such cycles $k$-cycles for brevity. 
If one of the intervals in (\ref{ee9}) is empty then only a single 
cycle exists (Figure~\ref{fig:clipped-2-cycle}). Otherwise, two 
cycles exist simultaneously (Figure~\ref{fig:1-2-cycles}); their 
orders are two subsequent positive integers. According to 
Theorem~\ref{prop_stab}, which cycle is realized depends on the 
initial conditions. 
 
\section{Properties of cycles} 
\label{sec:cycles} 
 
In this section, we characterize the shape of cycles. In other 
words, for given parameters $\beta$, $q$ and $b$, we would like to 
know if the limit cycles of the system trajectories are clipped or 
unclipped and what orders the cycles have. For fixed values of 
$\beta$ and $q$, we define the following quantities: 
\begin{equation}\label{ee1} 
N\defi\min\left\{i\ge 1:~\frac{\beta^i}{1-\beta^i}<q\right\}; 
\end{equation} 
\begin{equation}\label{ee2} 
D\defi \ln(1-\beta^N) + \frac{2\beta^N}{1-\beta^N}; 
\end{equation} 
\begin{equation}\label{ee3} 
C\defi-\ln(1-\beta^N)-\beta^N; 
\end{equation} 
$\theta_k$ is the single positive solution to equation 
\begin{equation}\label{ee4} 
\ln\frac{\theta}{1-e^{-\theta}}+\frac{\beta^k\theta}{1-\beta^k}=q-\frac{\beta^k}{1-\beta^k},~~~~~k=N,N+1; 
\end{equation} 
\begin{equation}\label{ee5} 
b_{0,k}\defi\frac{\theta_k}{1-e^{-\theta_k}}-\ln\frac{\theta_k}{1-e^{-\theta_k}}-1. 
\end{equation} 
Then, we define the set of quantities which do not depend on $q$:\\ 
$\tau_k$ is the single positive solution to equation 
\begin{equation}\label{ee6} 
\frac{\tau}{1+\frac{\beta^{k-1}-\beta^k}{1-\beta^k}(\tau+1)}=1-e^{-\tau},~~~k=2,3,\ldots 
\end{equation} 
\begin{equation}\label{ee7} 
A^*_k\defi\frac{\beta^{k-1}(\tau_k+1)}{1-\beta^k}; 
\end{equation} 
\begin{equation}\label{ee8} 
q^*_k\defi\frac{\beta^k}{1-\beta^k}(\tau_k+1)+\ln\frac{\tau_k}{1-e^{-\tau_k}}; 
\end{equation} 
It is convenient to put $\tau_1, A^*_1$ and $q^*_1$ equal to 
$+\infty$. Finally, in case $q \le D$ one has to solve equation 
   \begin{equation}\label{ee18} 
e^{-r}+r-1=\beta^N(q+r+1)-q. 
  \end{equation} 
It has no more than two positive solutions $\underline{r} \le \bar{r}$ which 
define 
  \begin{equation}\label{ee19} 
\underline{b}\defi e^{-\underline{r}}+\underline{r}-1; \quad 
\bar{b}\defi e^{-\bar{r}}+\bar{r}-1, 
  \end{equation} 
Note that $\underline{b} \le \bar{b}$. If $q \le q^*_{N+1}$ then $q \le D$ and 
$\bar{b} \ge A^*_{N+1}-q$. 
 
We note that all the above defined quantities do not depend on $b$. 
Thus, from now on we assume that $\beta$ and $q$ are fixed and we 
are going to describe what kind of cycles exist for different values 
of $b$. In other words, we study what effect the router buffer size has on the 
limiting behavior of TCP/IP. There are three cases: 
 
\medskip 
 
{\bf Case $A^*_{N+1}<q$.} 
 
If $b\in\left[0,\frac{\beta^{N-1}}{1-\beta^{N-1}}-q\right]$ then only the cycle 
of order $N$ exists. In case $N=1$, we put $\frac{\beta^0}{1-\beta^0}=+\infty$ for generality. 
 
Suppose $N>1$. Then for 
$b\in\left(\frac{\beta^{N-1}}{1-\beta^{N-1}}-q,A^*_N-q\right]$ two 
cycles, of orders $N$ and $N-1$ exist simultaneously. For 
$b\in\left(A^*_N-q,\frac{\beta^{N-2}}{1-\beta^{N-2}}-q\right]$, 
there exists only a single cycle of order $N-1$. And so on; for 
$b>A_2^*-q$, only 1-cycle exists (see Figure~\ref{fig3}). 
 
The $N$-cycle is clipped for $b\in[0,b_{0,N})$. Cycles of lower 
orders are unclipped for all values of $b$, if they exist. 
 
The $N$-cycle touches the $v$-axis at a single point iff 
$b=b_{0,N}$. Thus, if $b=b_{0,N}$ there exists a critical $N$-cycle. 
No critical cycles of lower orders exist. 
 
\begin{figure}[h] 
  \begin{minipage}{.5\textwidth} 
    \begin{center} 
        \centering {\epsfxsize=2.6in 
\epsfbox{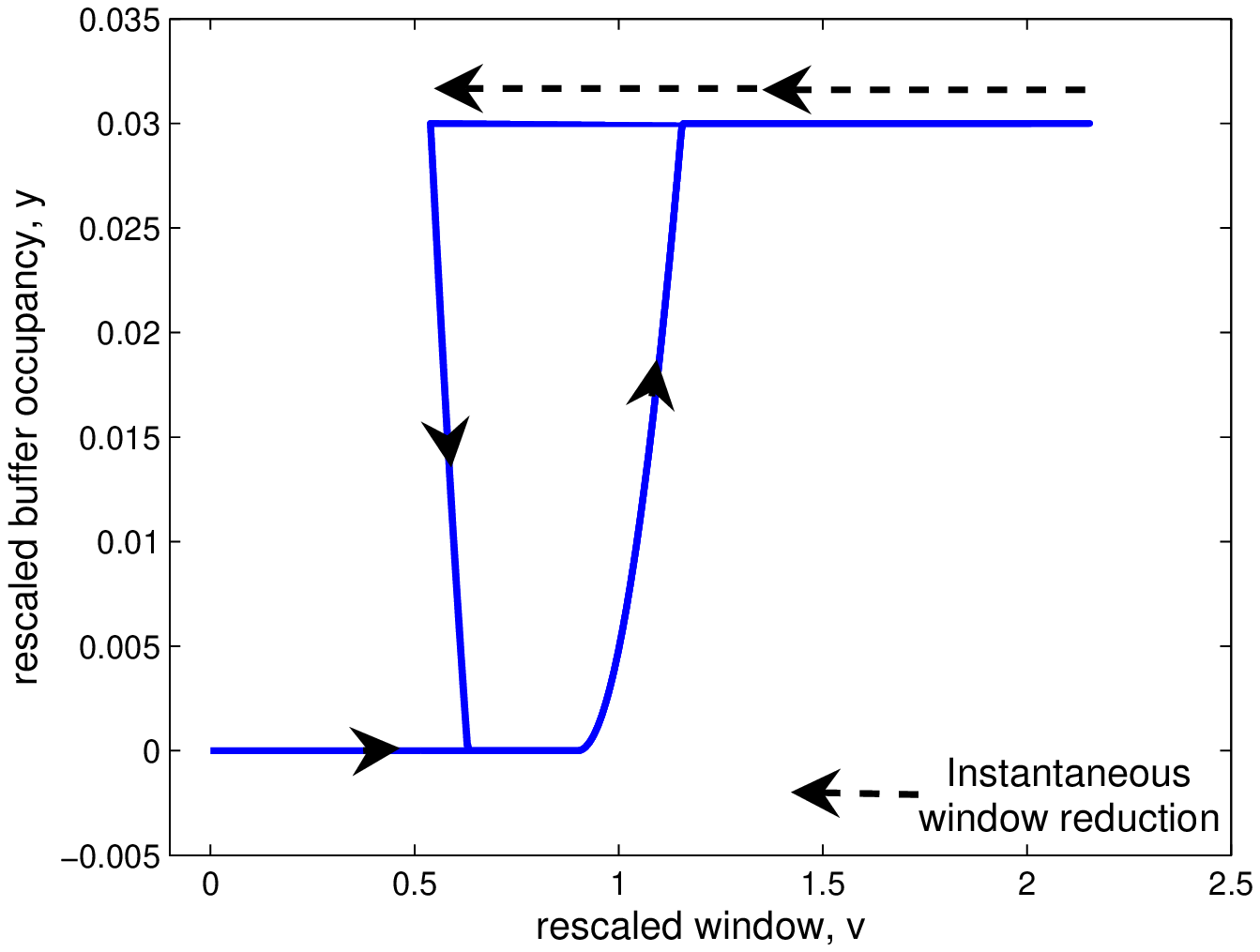}} 
        \caption{Clipped 2-cycle. Case $A_2^*<q$.} 
        \label{fig:clipped-2-cycle} 
    \end{center} 
  \end{minipage} 
\nolinebreak 
  \begin{minipage}{.5\textwidth} 
    \begin{center} 
\centering {\epsfxsize=2.6in 
\epsfbox{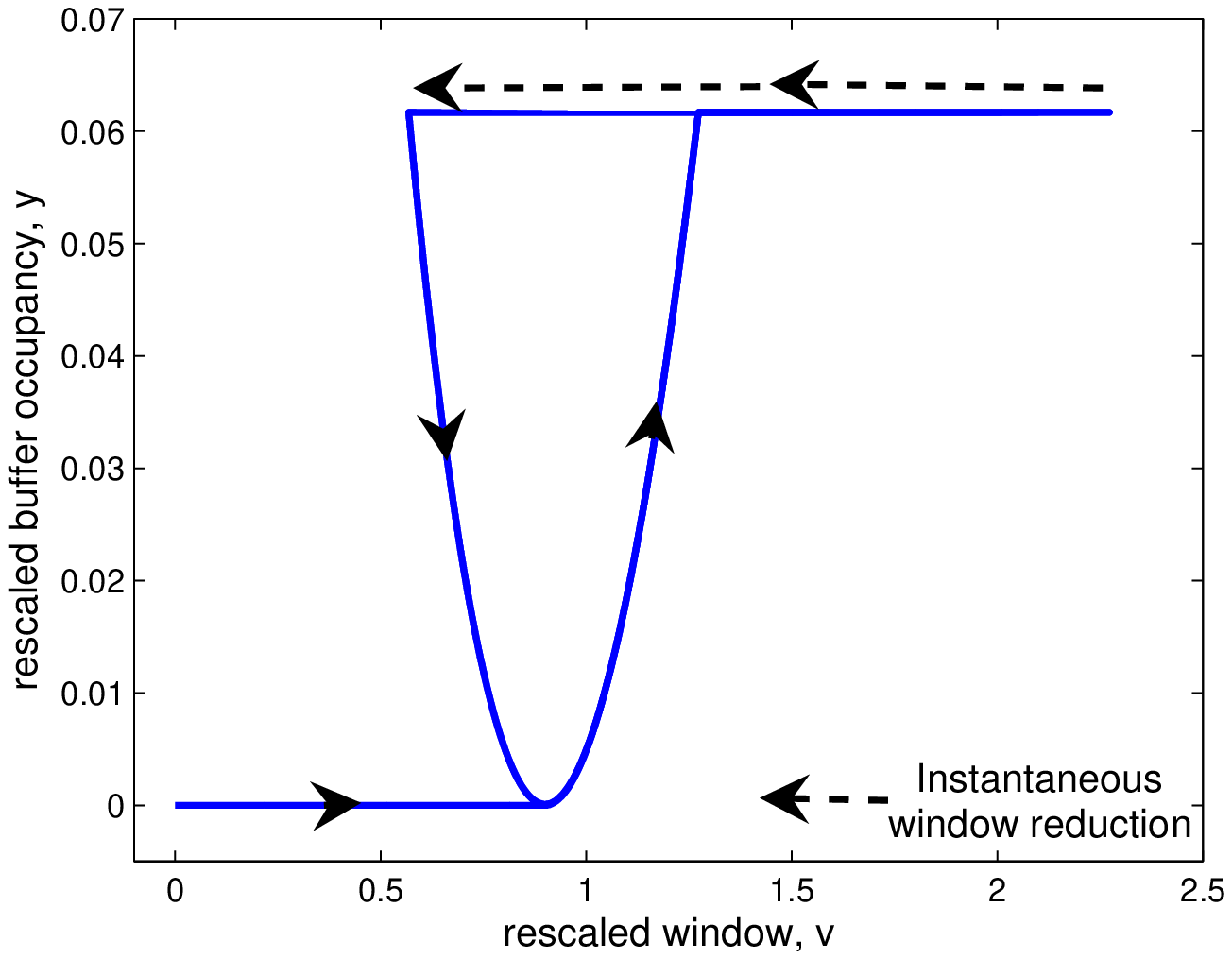}} 
\caption{Critical 2-cycle. Case $A_2^*<q$.} 
\label{fig:critical-2-cycle} 
    \end{center} 
  \end{minipage} 
\end{figure}

\begin{figure}[h] 
  \begin{minipage}{.333\textwidth} 
    \begin{center} 
        \centering {\epsfxsize=2.0in \epsfbox{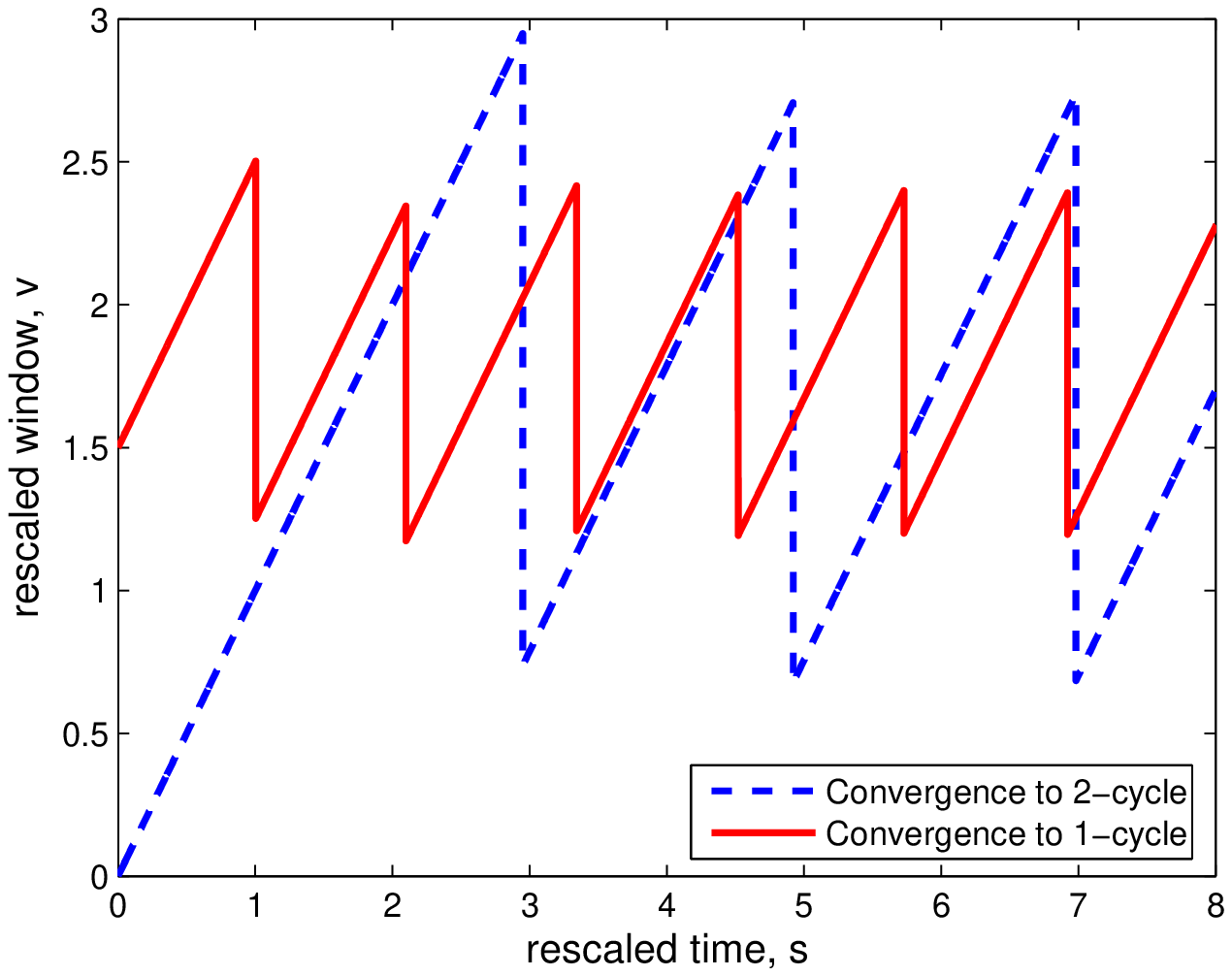}} 
        \label{fig:rate_time_a2smallerq} 
    \end{center} 
  \end{minipage} 
\nolinebreak 
  \begin{minipage}{.333\textwidth} 
    \begin{center} 
\centering {\epsfxsize=2.0in \epsfbox{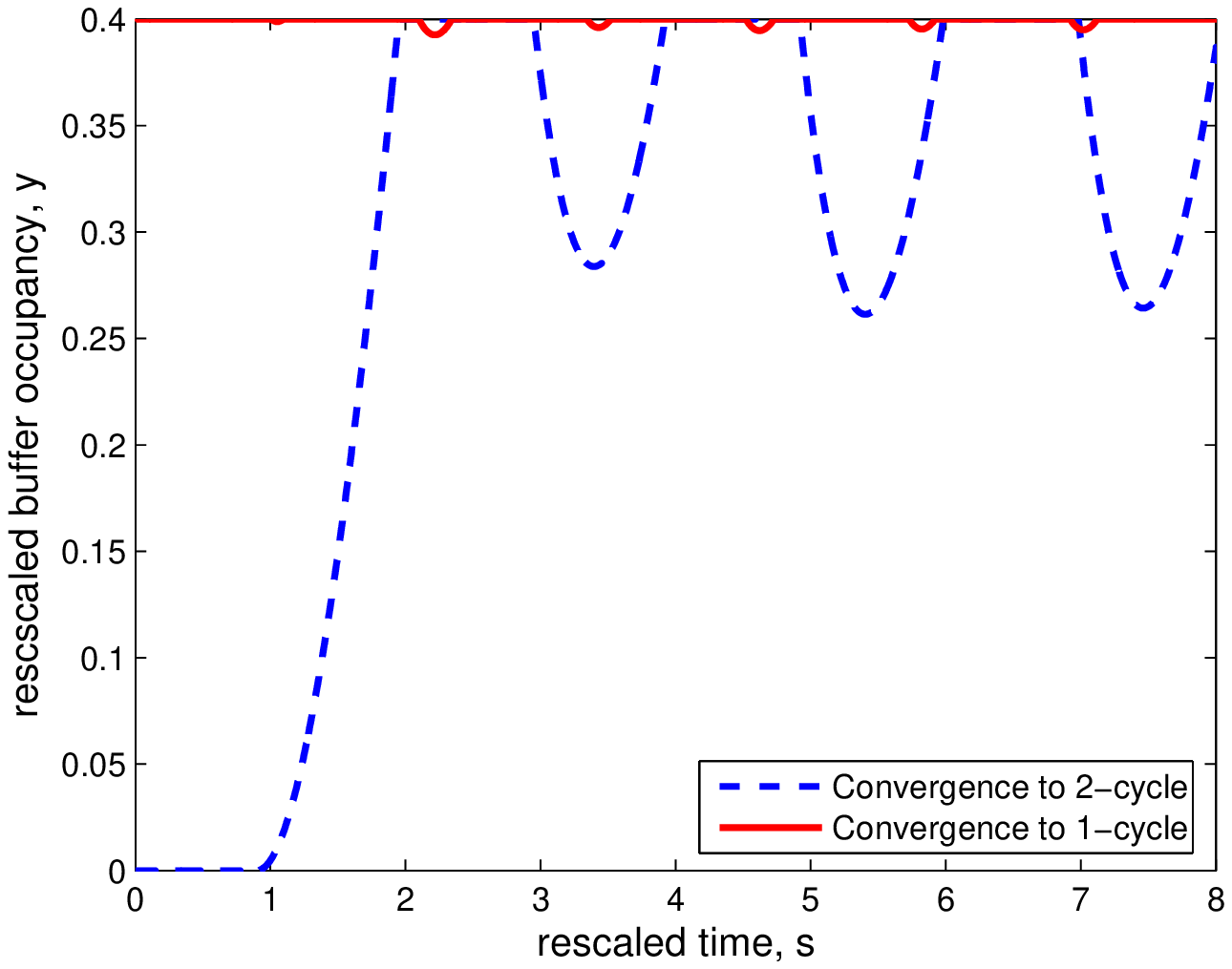}} 
\label{fig:buffer_time_a2smallerq} 
    \end{center} 
  \end{minipage} 
\nolinebreak 
  \begin{minipage}{.333\textwidth} 
    \begin{center} 
\centering {\epsfxsize=2.0in 
\epsfbox{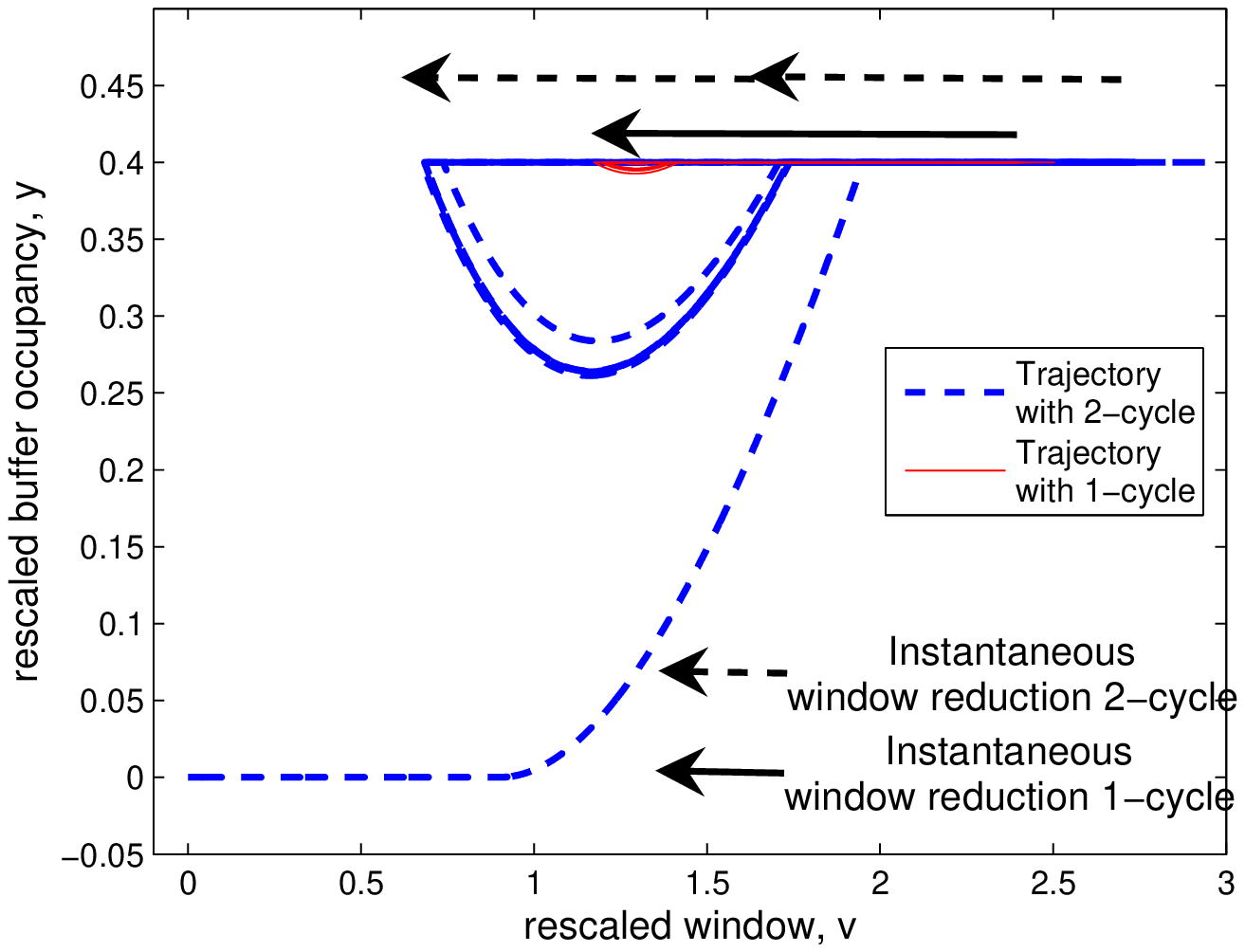}} 
\label{fig:rate_buffer_a2smallerq} 
    \end{center} 
  \end{minipage} 
\caption{Co-existence of 1-cycle and 2-cycle. Case $A_2^*<q$. } 
\label{fig:1-2-cycles} 
\end{figure}

\begin{example} 
Let us illustrate this with a numerical example. If we take $q=0.9$ 
and $\beta=1/2$ then $N=2$, $A_2^*=1.4965$, $A_3^*=0.3910$. If $b 
\in [0,0.1]$ we have only 2-cycles; if $b\in (0.1, 0.5965]$ we have 
1-cycles and 2-cycles $($see Figure~\ref{fig:1-2-cycles}$)$; and if 
$b>0.5965$ we have only 1-cycles. For each $b<b_{0,2}=0.0617$, there 
exists only a clipped 2-cycle $($see 
Figure~\ref{fig:clipped-2-cycle}$)$. As one can see on 
Figure~\ref{fig:critical-2-cycle}, when $b=b_{0,2}=0.0617$, the 
2-cycle becomes critical. All figures for this example have been 
plotted with MATLAB Simulink. 
\end{example} 
 
\medskip 
 
{\bf Case $q \le q^*_{N+1}$.} 
 
If $b \in [0,\underline{b})$, then only the $N$-cycle exists. If $b 
\in [\underline{b}, A^*_{N+1}-q]$, then two cycles of orders $N$ and 
$N+1$ exist simultaneously. For $b \in (A^*_{N+1}-q, 
\frac{\beta^{N-1}}{1-\beta^{N-1}}-q]$, again, only the $N$-cycle 
exists. 
 
 
The $N$-cycle is clipped for $b \in [0,b_{0,N})$; the $(N+1)$-cycle 
is clipped for $b \in [\underline{b},b_{0,N+1})$. These cycles 
become critical at $b=b_{0,N}$ and $b=b_{0,N+1}$, respectively. 
Cycles of lower orders are unclipped for all values of $b$, if they 
exist. If $N>1$ then, similarly to the case $A^*_{N+1}<q$, the order 
of the cycle decreases as $b$ increases above 
$\frac{\beta^{N-1}}{1-\beta^{N-1}}-q$ (see Figure~\ref{fig3}). 
 
\medskip 
 
{\bf Case $q^*_{N+1}<q\le A^*_{N+1}$.} 
 
If $C \le A^*_{N+1}-q$\ \ \footnote{Actually $C$ cannot be equal to 
$A^*_{N+1}-q$.} and $q \le D$, then everything is similar to the 
case $q\le q^*_{N+1}$. The difference is that the $(N+1)$-cycle is 
clipped and cannot be critical; it exists simultaneously with the 
$N$-cycle for $b\in[\underline{b},\bar b]$. If $b\in\left(\bar 
b,\frac{\beta^{N-1}}{1-\beta^{N-1}}-q\right]$, only the $N$-cycle 
exists. The latter interval is non-empty. 
 
If $C > A^*_{N+1}-q$ or $D < q$, then everything is exactly as in case 
$A^*_{N+1} < q$.

\section{Conditions for the absence of multiple jumps} 
\label{sec:multijumps} 
 
The regime with multiple jumps is not desirable. The multiple jump 
corresponds to the lost of more than one packet in a single 
congestion window. Subsequent packet losses can force TCP to switch 
from the Congestion Avoidance TCP phase to the Slow Start phase and 
lead to lengthy timeouts. Furthermore, the absence of subsequent 
packet losses is beneficial not only for the TCP performance but 
also for the quality of service provided to the end users. In the 
next theorem we provide necessary and sufficient conditions for the 
absence of multiple jumps, namely, we characterize all possible 
cases when only a single cycle of order $1$ exists. 
 
\begin{theor} \label{pro2} 
The following mutually exclusive conditions fully characterise all 
possible cases when only a single cycle of order $1$ exists: 
 
\begin{itemize} 
 
\item[(a)] $\frac{\beta}{1-\beta}\ge q$ and $b+q>A^*_2$; 
 
\item[(b)] $A^*_2<q$ ($b$ can be arbitrary); 
 
\item[(c)] $\frac{\beta}{1-\beta}<q\le q^*_2$ and $b\notin [\underline{b}, A^*_2-q]$; 
 
\item[(d)] $\max\left\{\frac{\beta}{1-\beta},~q^*_2\right\}<q \le A^*_2-C$, $q \le D$ 
and $b\notin [\underline{b},\bar b]$; 
 
\item[(e)] $\max\left\{\frac{\beta}{1-\beta},~q^*_2\right\}<q \le A^*_2-C$, $q > D$ 
($b$ can be arbitrary); 
 
\item[(f)] $\max\left\{\frac{\beta}{1-\beta},~q^*_2,~A^*_2-C\right\}<q \le A^*_2$ 
($b$ can be arbitrary). 
 
\end{itemize} 
 
\end{theor} 
 
In the following corollary we provide a simple sufficient condition 
for absence of multiple jumps. 
 
\begin{conse} 
Condition $b+q>A^*_2$ is sufficient for the absence of cycles of 
orders $k>1$. (See Figure~\ref{fig3} and Corollary \ref{co4}.) 
\end{conse} 
 
\begin{figure}[h] 
  \begin{minipage}{.5\textwidth} 
    \begin{center} 
        \centering {\epsfxsize=2.6in \epsfbox{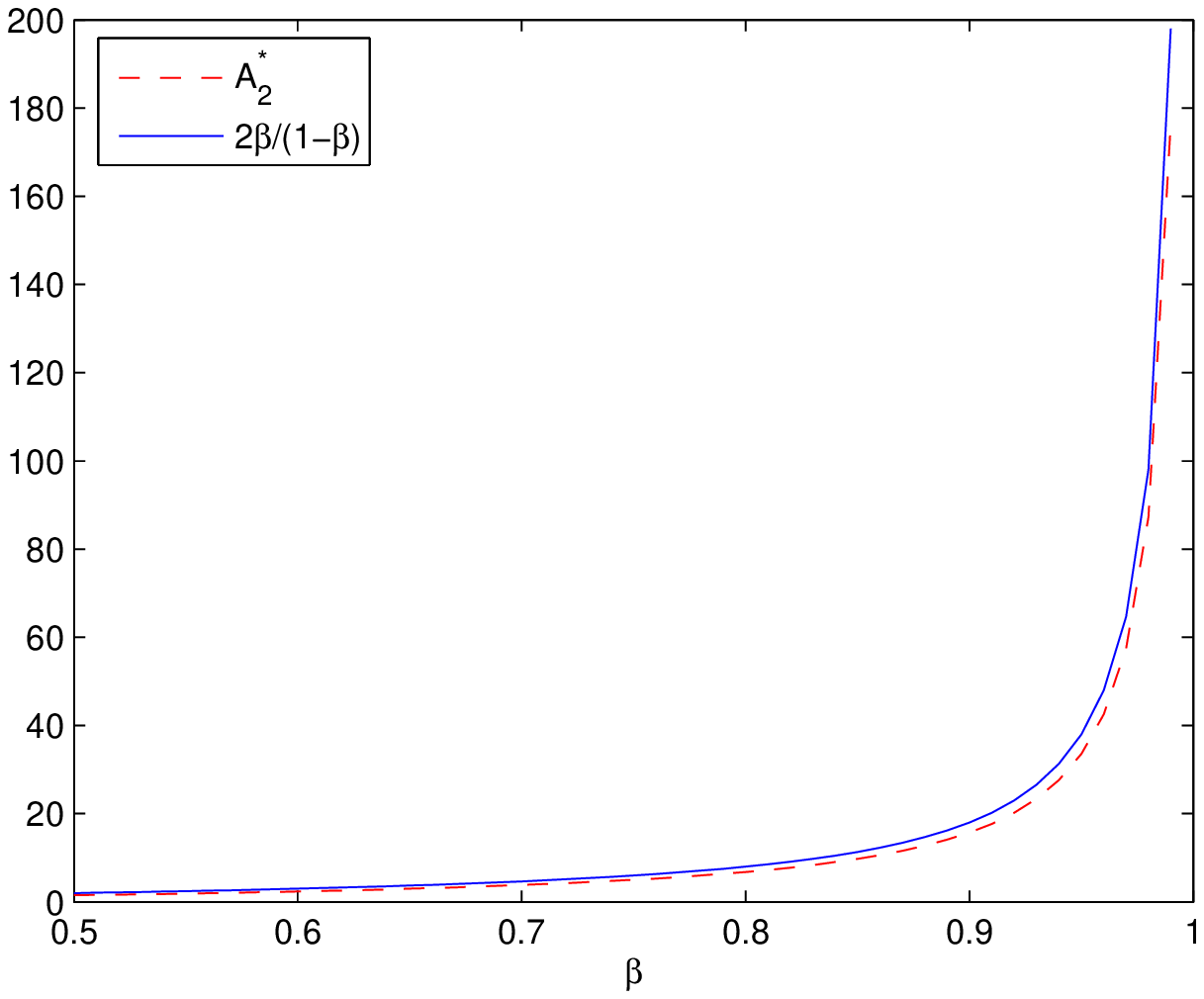}} 
        \caption{Comparison of sufficient conditions for the absence of multiple jumps.} 
        \label{fig:absence_multiple_jumps_1} 
    \end{center} 
  \end{minipage} 
\nolinebreak 
  \begin{minipage}{.5\textwidth} 
    \begin{center} 
\centering {\epsfxsize=2.6in \epsfbox{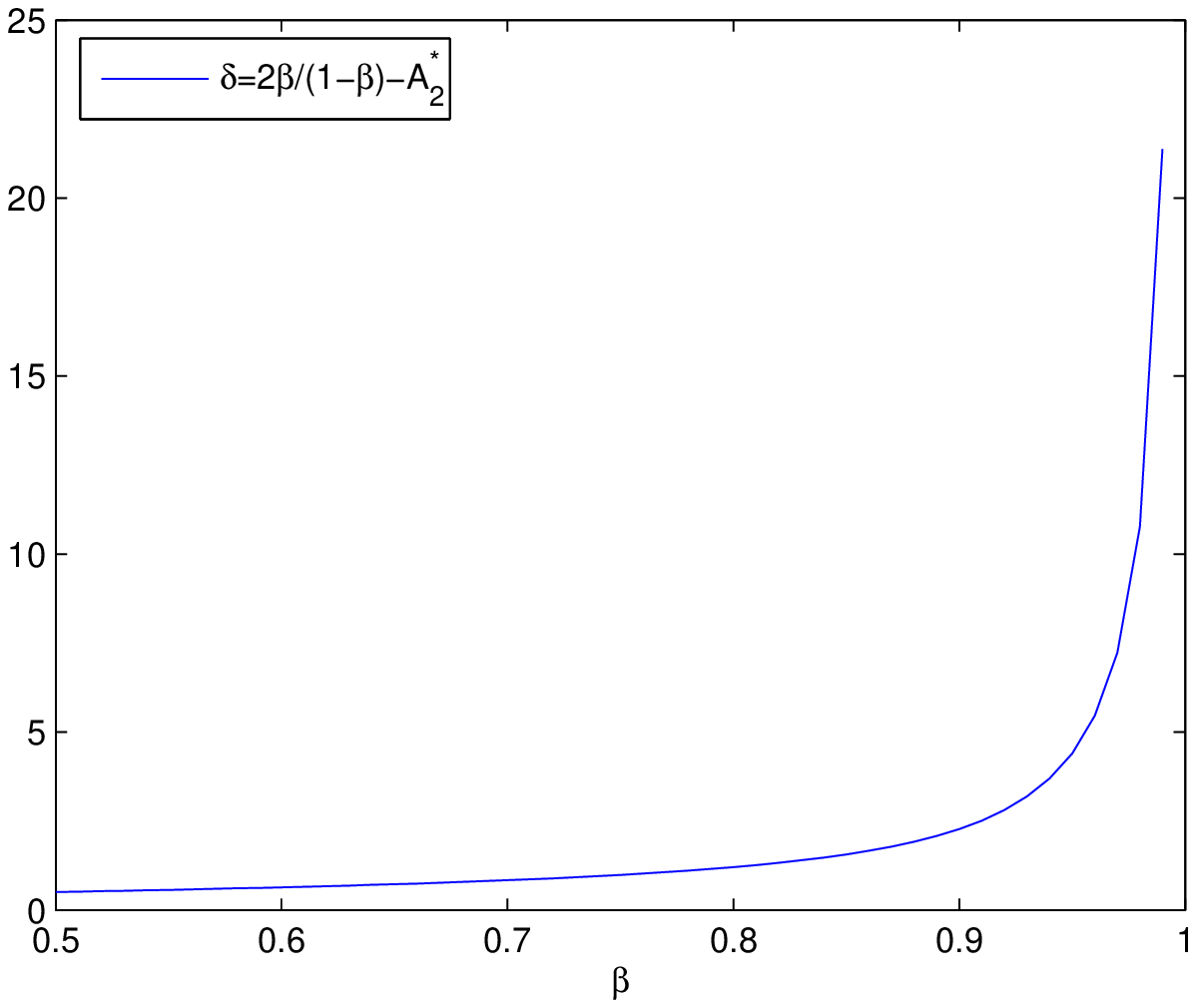}} 
\caption{The value of $2\beta/(1-\beta)-A^*_2(\beta)$.} 
\label{fig:absence_multiple_jumps_2} 
    \end{center} 
  \end{minipage} 
\end{figure} 
 
Recall that $A^*_2$ depends only on $\beta$. In particular, if 
$\beta=1/2$, $A^*_2=1.4965$ . 
 
We would like to note that the above sufficient condition is tighter 
than the sufficient condition for the absence of multiple jumps 
provided in \cite{HBOL01}: 
$b+q > 2\beta/(1-\beta)$. To compare these two 
conditions, we plot $A^*_2(\beta)$ and $2\beta/(1-\beta)$ 
in Figure~\ref{fig:absence_multiple_jumps_1}  and the difference 
$2\beta/(1-\beta)-A^*_2(\beta)$ in 
Figure~\ref{fig:absence_multiple_jumps_2}. Strictly speaking we have 
 
\begin{proposition} \label{pro3} 
The difference $\delta\defi\frac{2\beta}{1-\beta}-A^*_2$ is always 
positive and $\lim_{\beta\to 1}\delta =+\infty$. 
\end{proposition} 
 
Nevertheless, the simple sufficient condition of \cite{HBOL01} 
appears to be quite good except for values of $\beta$ that are too 
close to one.

\section{Pareto set for optimal buffer sizing} 
\label{sec:pareto} 
 
\begin{figure}[h] 
  \begin{minipage}{.5\textwidth} 
    \begin{center} 
\centering {\epsfxsize=2.6in \epsfbox{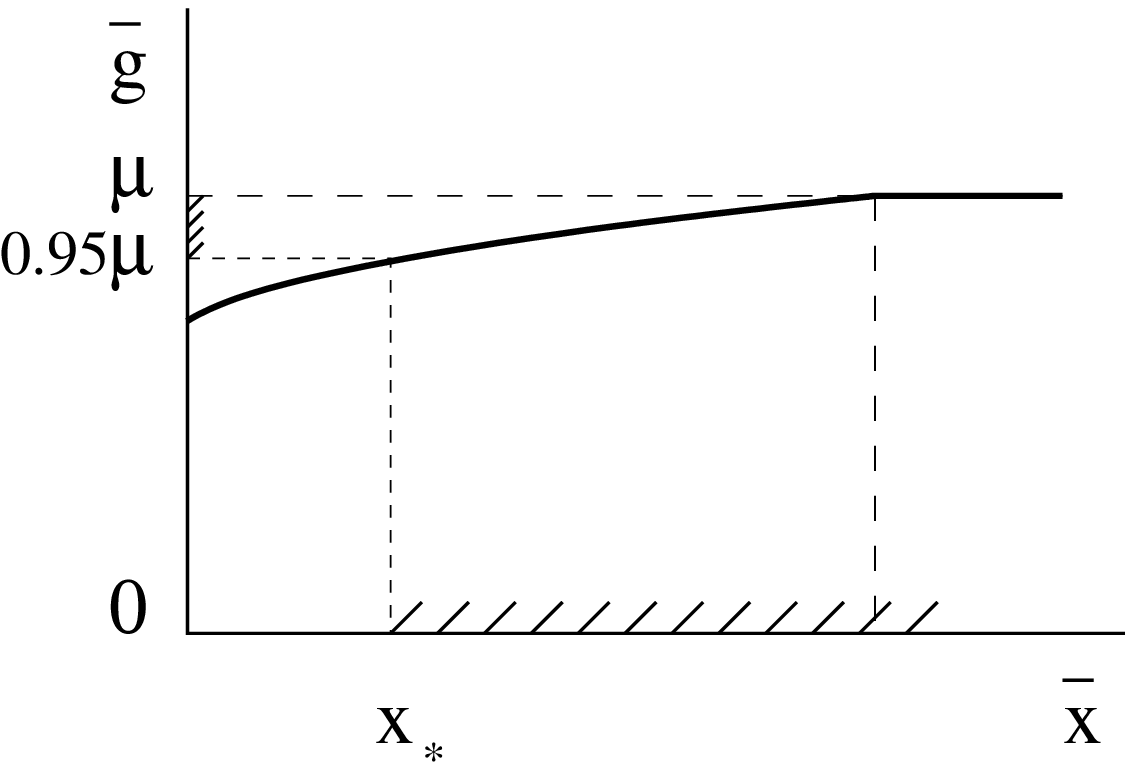}} 
\caption{Pareto set.} \label{fig:Pareto_ex} 
    \end{center} 
  \end{minipage} 
\nolinebreak 
    \begin{minipage}{.5\textwidth} 
    \begin{center} 
        \centering {\epsfxsize=2.6in \epsfbox{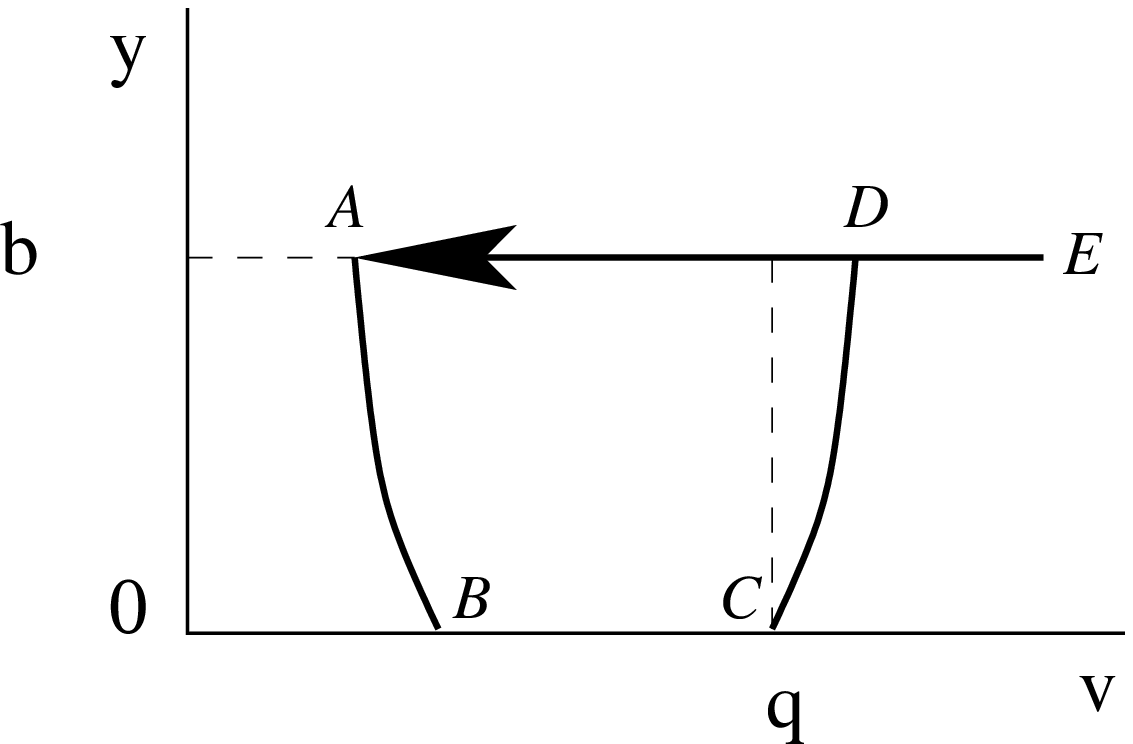}} 
        \caption{Phases of the clipped cycle.} 
        \label{fig:ClipCycle} 
    \end{center} 
  \end{minipage} 
\end{figure}

Let us study what effect has the choice of the buffer size on the 
performance of TCP. In particular, we are interested in optimal 
buffer sizing. Towards this goal, let us formulate the performance 
criteria. On one hand, we are interested to obtain as large goodput 
as possible. That is, we are interested to maximize the average 
goodput 
$$ 
\bar{g} = \lim_{t \to \infty} \frac{1}{t} \int_0^t g(s) ds, 
$$ 
where the instantaneous goodput $g(t)$ is defined by 
$$ 
g(t)=\left\{\begin{array}{ll} \lambda(t), & \mbox{if} \quad x(t) < B,\\ 
\mu, & \mbox{if} \quad x(t)=B. 
\end{array}\right. 
$$ 
On the other hand, we are interested to make the delay of data in 
the buffer as small as possible. That is, we are also interested to 
minimize the average amount of data in the buffer 
$$ 
\bar{x} = \lim_{t \to \infty} \frac{1}{t} \int_0^t x(s) ds. 
$$ 
Clearly, these two goals are contradictory. In fact, here we face a 
typical example of multi-criteria optimization. A standard approach to it 
is to consider the optimization of 
one criterion under constraints for the other criteria (see e.g., 
\cite{P97}). Namely, we would like to maximize the goodput given 
that the average amount of data in the buffer does not exceed a 
certain value 
\begin{equation} 
\label{max_rate} \max\{ \bar{g} : \bar{x} \le \bar{x}_* \}. 
\end{equation} 
Or we would like to minimize the average delay given that the 
average goodput is not less than a certain value 
\begin{equation}f 
\label{min_delay} \min\{ \bar{x} : \bar{g} \ge \bar{g}_* \}. 
\end{equation} 
The solution to the above constrained optimization problems can be 
obtained from the Pareto set. As is known, see e.g. \cite{P97}, the 
Pareto set can be constructed by solving the optimization problem 
\begin{equation} 
\label{opt_uncon} \max \left\{ \lim_{t \to \infty} \frac{1}{t} 
\int_0^t c_1 g(s) - c_2 x(s) ds \right\}. 
\end{equation} 
To be more precise, the Pareto Set is formed by the pairs of 
objectives $(\bar{g},\bar{x})$ that solve (\ref{opt_uncon}) for 
different $(c_1,c_2) \in {R}^2_{+}$. An example of Pareto set is 
given in Figure~\ref{fig:Pareto_ex}. Each point of the Pareto set 
corresponds to a solution of optimization problem (\ref{opt_uncon}) 
for some choice of $c_1$ and $c_2$. Once we obtain the Pareto set, 
it is very easy to deduce solution of problems (\ref{max_rate}) and 
(\ref{min_delay}). For instance, if one wants that the utilization 
of the bottleneck router will be not less than, say, 95\%, one has 
to be ready to accept the delays that are equal or greater than 
$x_*$. 
 
All three optimization problems (\ref{max_rate}), (\ref{min_delay}) 
and (\ref{opt_uncon}) can be regarded as mathematical formulation of 
the lingual criterion ``find the link buffer size that accommodates 
both TCP and UDP traffic'' given in \cite{GKT04}. Since UDP traffic 
does not contribute much in terms of the load, for the design of IP 
routers one can use for instance optimization problem 
(\ref{max_rate}) where the delay constraint is imposed by the UDP 
traffic. 
 
We note that here we deal with the optimal impulse control problem 
of a deterministic system with long-run average optimality 
criterion. To the best of our knowledge there are no available 
results on such type of problems in the literature. In principle, 
the control policy in our model can depend on the current values 
of $x$ and $\lambda$. In practice, however, all currently 
implemented buffer management 
schemes (e.g., AQM, DropTail) send congestion signals based only on 
the state of the buffer. Thus, we also limit ourselves to the case 
when the control depends only on the amount of data in the buffer. 
Furthermore, we restrict the control action only to the choice of 
the buffer size. Thus, the control signal is only sent at the moment 
when the buffer gets full. 
 
The following theorem provides expressions for the average sending 
rate, goodput and queue size under condition $q > A_2^*$, which 
guarantees the absence of multiple jumps for any value of the buffer 
size. Remember that $A^*_2$ depends only on $\beta$ (see 
(\ref{ee6}),(\ref{ee7})). In particular, the expressions allow us to 
plot the Pareto set parameterized by the buffer size. 
 
\begin{theor} 
\label{th_pareto} Let the condition $\mu T/m > A_2^*$ be satisfied. 
Then, for $B \in [0,m b_{0,1}]$ the average sending rate, goodput 
and buffer occupancy are given by 
$$ 
\bar{\lambda} = \frac{m(1-\beta^2)}{2T_{cycle}}\left(1+\frac{\mu 
T}{m}+S_{CD}\right)^2, 
$$ 
$$ 
\bar{g}= \frac{m}{T_{cycle}} \left[\frac{1}{2}\left(\frac{\mu 
T}{m}+S_{CD}\right)^2-\frac{\beta^2}{2}\left(1+\frac{\mu 
T}{m}+S_{CD}\right)^2+\frac{\mu T+B}{m}\right], 
$$ 
$$ 
\bar{x}=\frac{1}{T_{cycle}} 
\left[mT\left(\int_0^{S_{AB}}y_{AB}(s)ds+\int_0^{S_{CD}}y_{CD}(u)du\right)\right. 
$$ 
$$ 
\left.+\frac{m^2}{\mu}\left(\int_0^{S_{AB}}y^2_{AB}(s)ds+\int_0^{S_{CD}}y^2_{CD}(u)du 
+\frac{B(\mu T+B)}{m^2}\right)\right], 
$$ 
respectively, where $T_{cycle}$ is the cycle duration given by 
$$ 
T_{cycle}  =  (1-\beta)(1+\frac{\mu T}{m}+S_{CD})T 
+\frac{B}{\mu}+\frac{m}{\mu}\left(\int_0^{S_{AB}}y_{AB}(s)ds 
+\int_0^{S_{CD}}y_{CD}(u)du\right), 
$$ 
with 
$$ 
y_{CD}(u)=e^{-u}+(u-1), 
$$ 
$$ 
y_{AB}(s)=[\frac{B}{m}+(1-\beta)(1+\frac{\mu T}{m})-\beta 
S_{CD}]e^{-s} 
+(s-1)+\beta(S_{CD}+1)-(1-\beta)\frac{\mu T}{m}, 
$$ 
where $S_{CD}$ and $S_{AB}$ are the solutions of the equations 
$$ 
e^{-S_{CD}}+S_{CD}-1=\frac{B}{m}, 
$$ 
$$ 
\left[\frac{B}{m}-\beta S_{CD}+(1-\beta)(1+\frac{\mu 
T}{m})\right]e^{-S_{AB}}+S_{AB} 
+\beta S_{CD} -(1-\beta)(1+\frac{\mu T}{m}) = 0. 
$$ 
For $B \in (m b_{0,1},\infty)$, we have 
$$ 
\bar{\lambda}=\frac{m}{2T_{cycle}}\frac{1+\beta}{1-\beta}(s_1+1)^2, 
$$ 
$$ 
\bar{g}=\mu, 
$$ 
$$ 
\bar{x}=\frac{1}{T_{cycle}} \left[mT\int_0^{s_1}y(s)ds 
+\frac{m^2}{\mu}\left(\int_0^{s_1}y^2(s)ds+\frac{B(\mu 
T+B)}{m^2}\right)\right], 
$$ 
where 
$$ 
T_{cycle}=T(s_1+1)+\frac{m}{\mu}\left(\int_0^{s_1}y(s)ds+\frac{B}{m}\right) 
$$ 
with 
$$ 
y(s)=\left[1+\frac{\mu T + B}{m}-v_0\right]e^{-s} + (s-1) + 
v_0-\frac{\mu T}{m}, 
$$ 
where $v_0$ and $s_1$ are defined by $(\ref{ea5})$ and $(\ref{ea6})$ 
with $k=1$. 
\end{theor} 
 
\begin{figure}[h] 
  \begin{minipage}{.5\textwidth} 
    \begin{center} 
\centering {\epsfxsize=2.6in \epsfbox{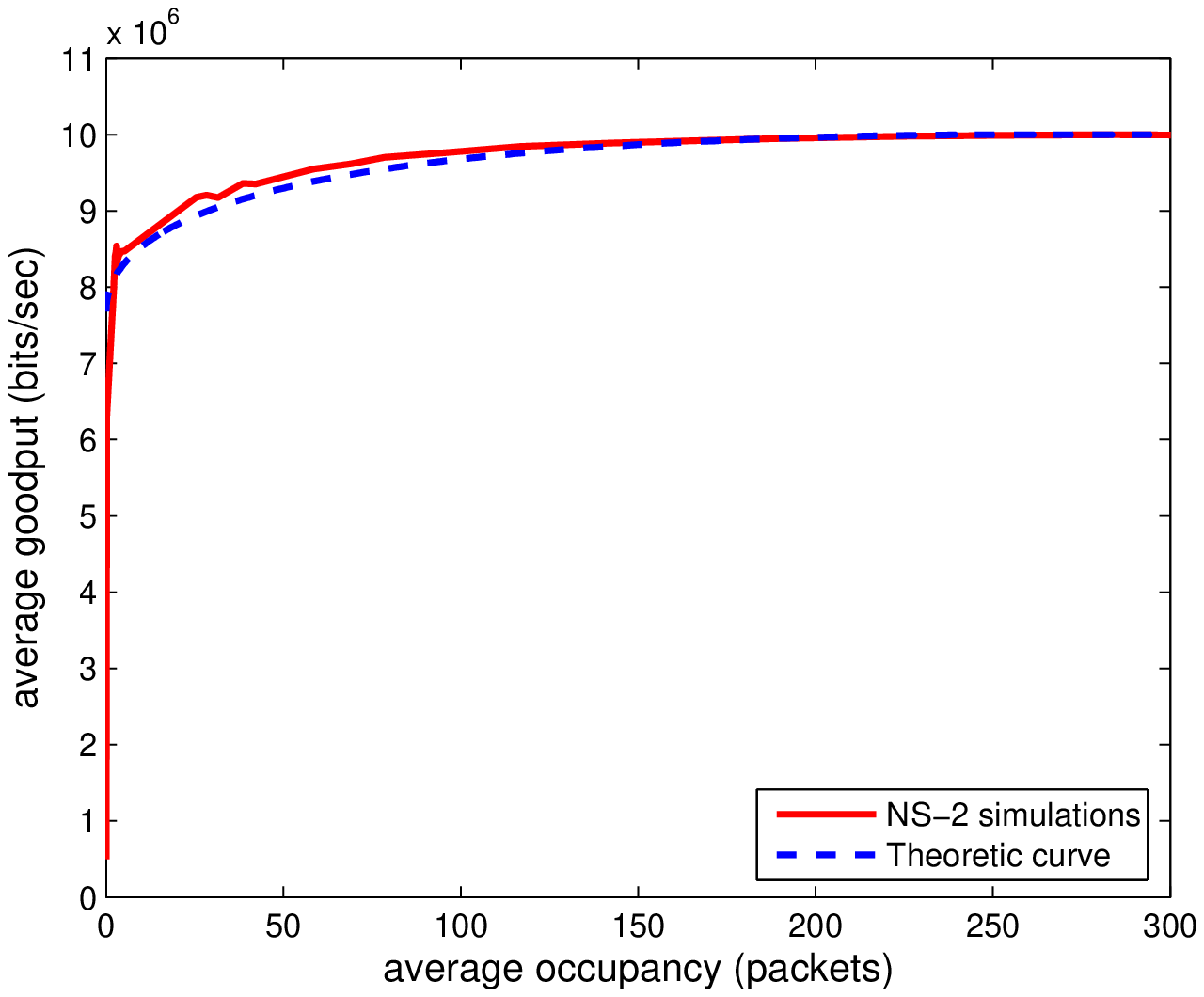}} 
\caption{Pareto set: Numeric calculations and NS-2 simulations.} 
\label{fig:paretoset_numerics} 
    \end{center} 
  \end{minipage} 
\nolinebreak 
    \begin{minipage}{.5\textwidth} 
    \begin{center} 
        \centering {\epsfxsize=2.6in \epsfbox{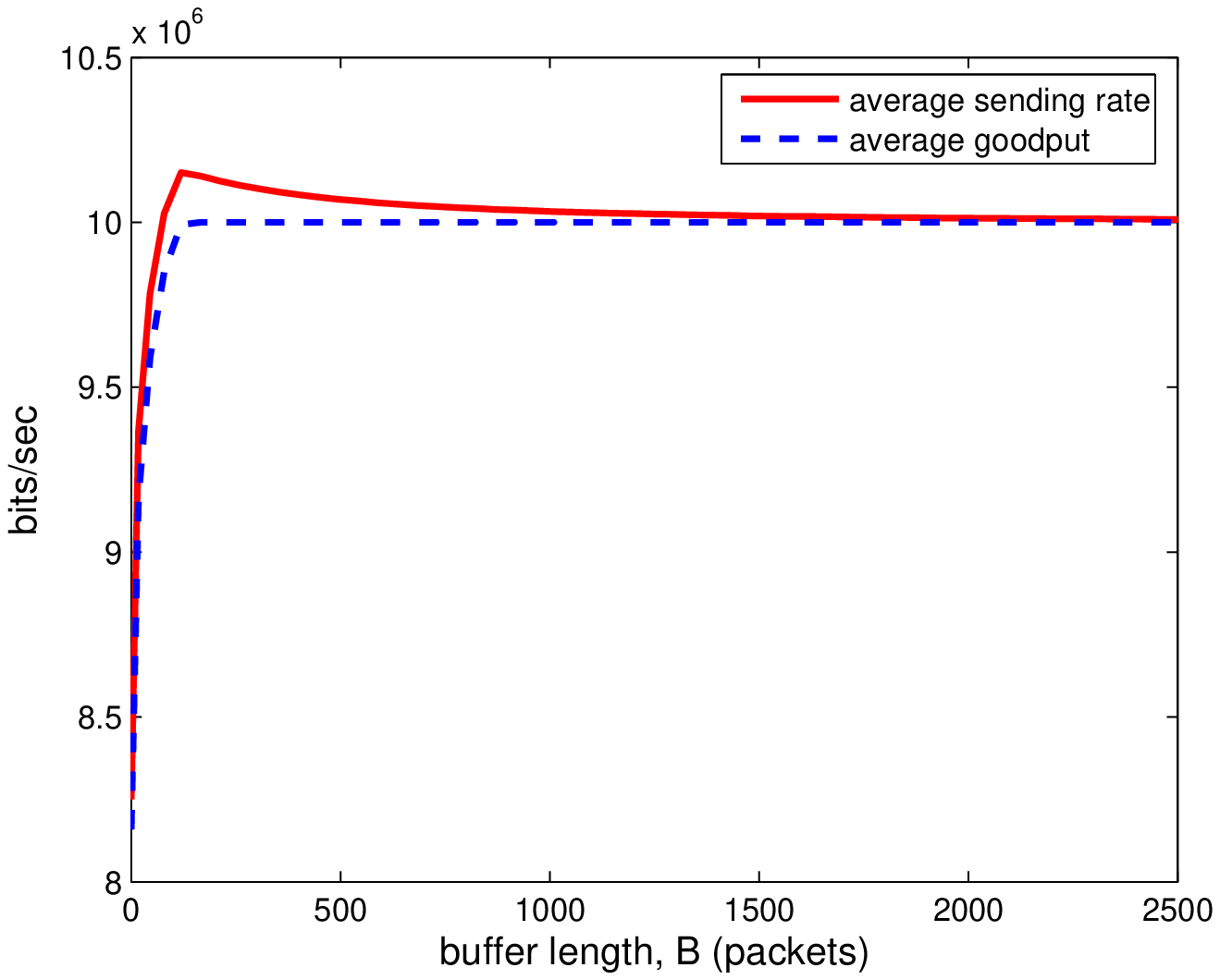}} 
        \caption{Non-monotonicity of the average sending rate.} 
        \label{fig:non_monotonicity_sendingrate} 
    \end{center} 
  \end{minipage} 
\end{figure}

\begin{example} 
\label{NSexample} Let us illustrate the Pareto set for a benchmark 
example of the TCP/IP network created with the help of NS-2 
simulator $\cite{NS}$. The network consists of a single bottleneck 
link of capacity $\mu=10Mbps$ which is shared by $n$ long-lived TCP 
connections. The propagation delay for each connection is $T=0.24s$
and $\beta=1/2$. The packet size is $4000bits$. Thus, we have that 
$m_0=4000bits$ as well. 
In Figure~\ref{fig:paretoset_numerics} we plot the Pareto set for 
$n=10$ (and $m=nm_0=40,000$) using the formulae of Theorem~\ref{th_pareto} and 
measurements obtained from NS simulations. As one can see, two 
curves match well. In Figure~\ref{fig:non_monotonicity_sendingrate}, 
again using the formulae of Theorem~\ref{th_pareto}, we plot the 
average goodput and the average sending rate as functions of the 
buffer size for $n=60$. 
\end{example} 
 
We note that $\bar{g}\le\mu$ always, but the average sending rate 
$\bar{\lambda}$ can exceed the router capacity $\mu$ (see 
Figure~\ref{fig:non_monotonicity_sendingrate}). Nevertheless, as the 
next Proposition~\ref{pro5} states, the difference between the 
average sending rate and the router capacity goes to zero as $B$ 
increases. In particular, this means that when the Drop Tail router 
is used, the rate of lost (and then retransmitted) information 
eventually diminishes to zero as the buffer size increases. 
 
\begin{proposition} 
\label{pro5} When $B \to \infty$, the difference 
$\Delta=\bar{\lambda}-\mu$ approaches zero from above. 
\end{proposition} 
 
\section{Minimal buffer size for the full system utilization} 
\label{sec:bmin} 
 
In the case of multiple TCP connections competing for resource of 
the bottleneck router we have $m=n m_0$. Here $n$ is the number of 
competing TCP connections. Let us study how the minimal buffer size 
for the full system utilization, $B_{0,N}$, depends on $n$ or, equivalently, on 
$m$. $B_{0,N}$ is the buffer size corresponding to scenario 
when the Pareto set touches the level $\mu$ (see 
Figure~\ref{fig:Pareto_ex}). It corresponds also to the critical cycle of 
minimal order.  
 
\begin{proposition} 
\label{pro4} $(a)$ For a fixed $N$, the value of $B_{0,N}=mb_{0,N}$ 
decreases as $m$ increases.\\ 
$(b)$ The value of $B_{0,N}$ increases as $N$ increases. 
\end{proposition} 
 
\begin{conse} 
\label{co8} 
The buffer size $B_{0,N}$ of the minimal order critical 
cycle is a piece-wise differentiable function of $m$, decreasing on 
the intervals $[m_i,m_{i+1})$; 
  $$\lim_{m\to m_{i+1}-0} B_{0,N}(m) < B_{0,N}(m_{i+1}),~~~i=0,1,2,\ldots$$ 
Here $m_i\defi \mu T (1-\beta^i)/\beta^i$; the value of $N$ equals 
$i+1$ on the interval $[m_i, m_{i+1})$ $($see $($\ref{ee1}$)$ $)$. 
 
Moreover, $\lim_{m\to m_N-0} B_{0,N}=0$, $\lim_{m\to m_N-0} \frac{ 
dB_{0,N}}{dm}=0$, $\lim_{m\to 0+} B_{0,1}=\mu T(1-\beta)/\beta$, 
$\lim_{N\to\infty} m_{N-1} B_{0,N}(m_{N-1})=0.5(\mu T(1-\beta))^2$ 
and hence $\lim_{m\to\infty} B_{0,N}=0$. 
\end{conse} 
 
\begin{figure}[h] 
  \begin{minipage}{.5\textwidth} 
    \begin{center} 
\centering {\epsfxsize=3.0in \epsfbox{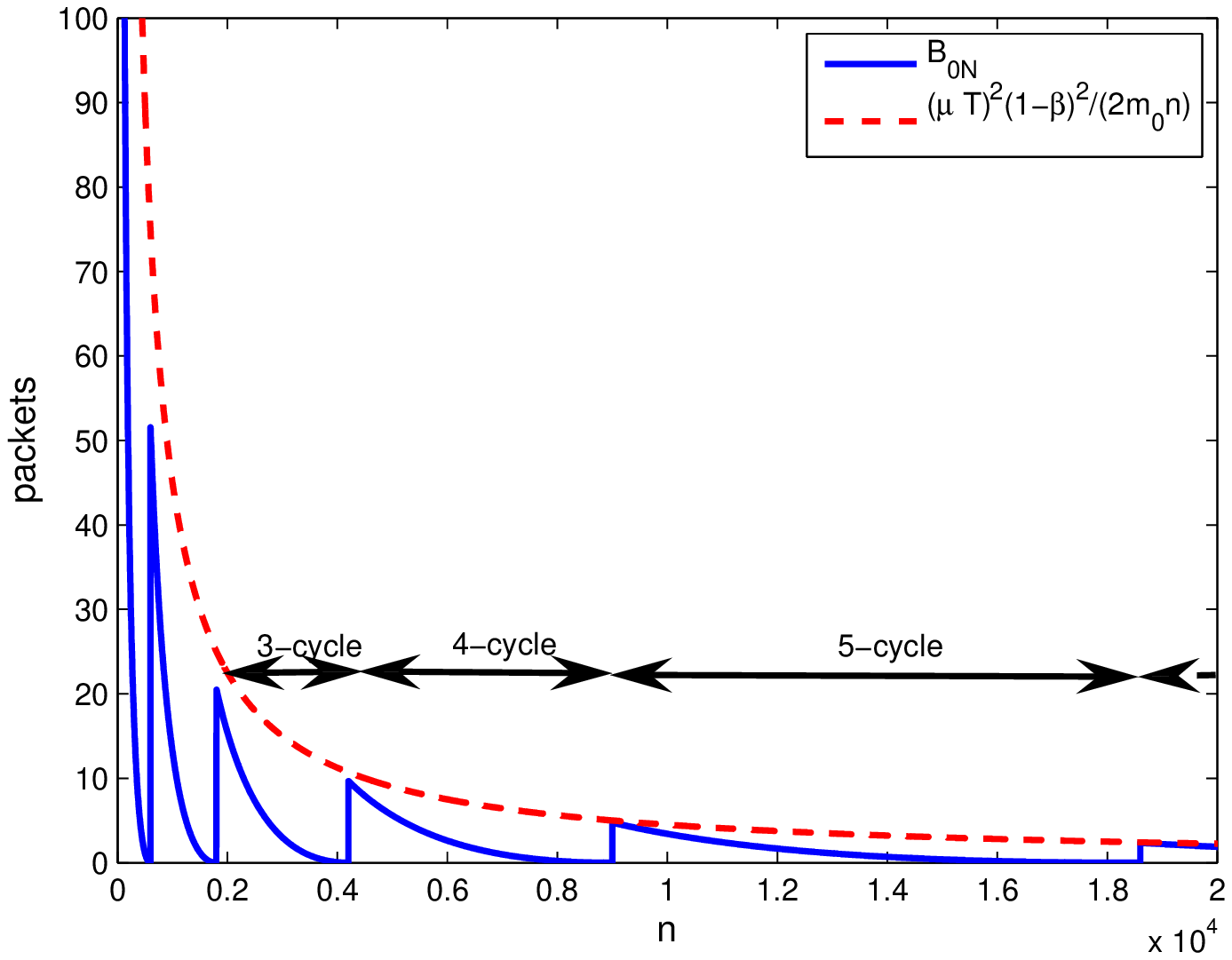}} 
\caption{The minimal buffer size for the full system utilization.} 
\label{fig:Bon} 
    \end{center} 
  \end{minipage} 
\nolinebreak 
    \begin{minipage}{.5\textwidth} 
    \begin{center} 
        \centering {\epsfxsize=3.0in \epsfbox{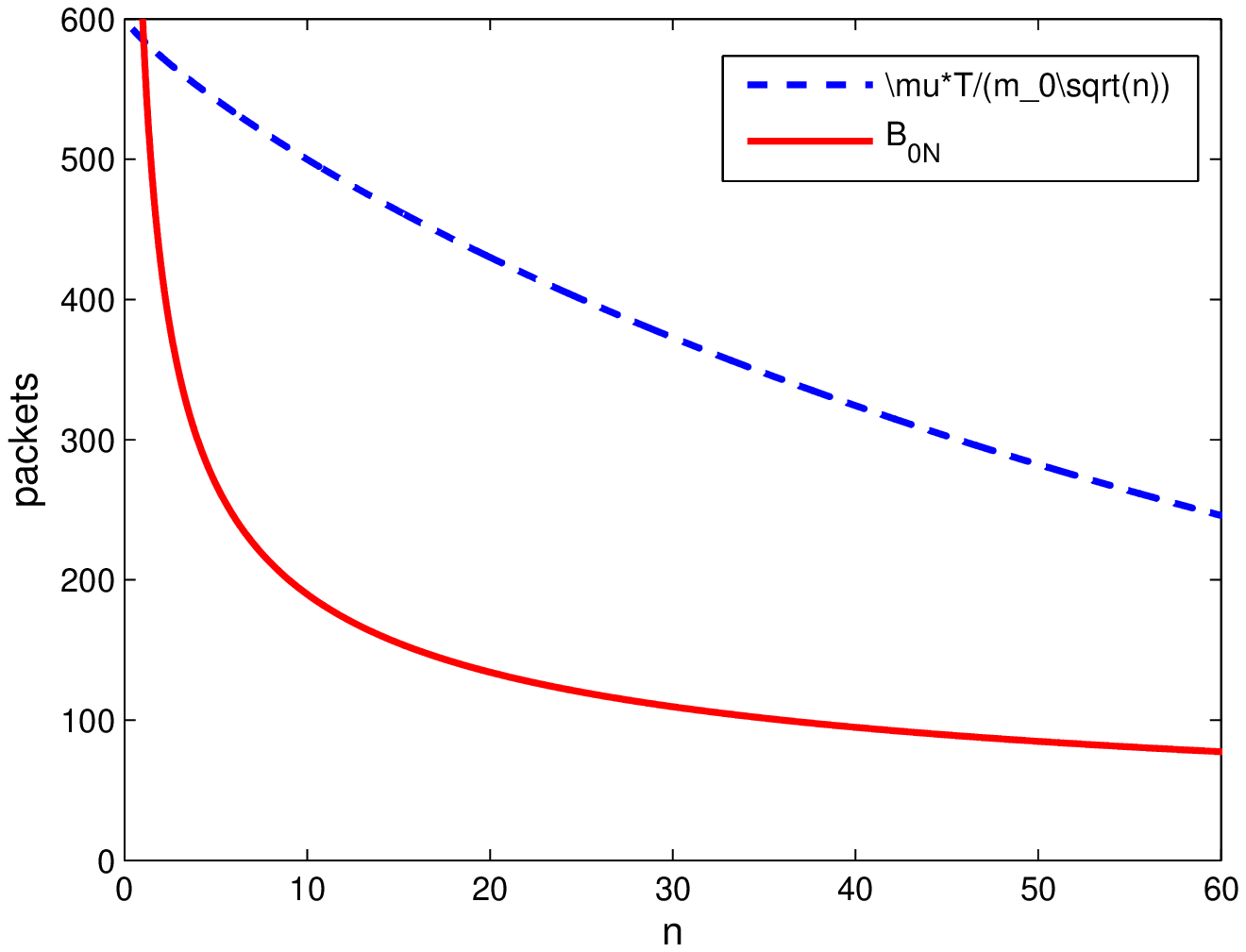}} 
        \caption{The minimal buffer size for the full system utilization (zoom).} 
        \label{fig:Bon_zoom} 
    \end{center} 
  \end{minipage} 
\end{figure}

\noindent {\bf Example~\ref{NSexample}(cntd.)} {\it In 
Figure~\ref{fig:Bon} we plot the buffer size $B_{0,N}$ of the 
minimal order critical cycle and the curve $f(m)=(1-\beta)^2(\mu 
T)^2/(2m)$ for $\mu T=2.4\times10^6bits \ (600 packets)$. The curve $f(m)$ 
indeed approaches fast the local maxima of $B_{0,N}$ as $m$ 
increases. In Figure~\ref{fig:Bon_zoom} we make a zoom on the 
interval with smaller values of $m$. As one can see, when $m$ goes 
to zero, the value of $B_{0,N}$ approaches 600packets, which is the 
BDP in this network example.} 
 
\bigskip 
 
We note that by Corollary~\ref{co8} for small values of $m$ the 
minimal buffer size for the full system utilization is approximately 
equal to $\mu T$, BDP of the bottleneck link. This is in agreement 
with the empirical conclusion of \cite{VS94}. In \cite{AKM04} the 
authors suggested that the minimal buffer size for the full system 
utilization should decrease as $(\mu T)/\sqrt{n}$ as the number of 
connections $n$ increases. We note that the authors of \cite{AKM04} 
have assumed that the competing TCP connections are not 
synchronized. That is, only a single connection reduces its 
congestion window when the buffer becomes full. In our model we 
assume full synchronization of competing TCP connections. Namely, 
when the buffer is full, all connections simultaneously reduce their 
congestion windows. We expect that the situation in real networks is 
in between these two extremes. And thus, the model of \cite{AKM04} 
provides an upper bound and our model provides a lower bound. 
Furthermore, it was believed previously that if the competing TCP 
connections are synchronized, one has to provide BDP of buffering to 
guarantee the full system utilization. From 
Figure~\ref{fig:Bon_zoom} one can see that the minimal buffer 
requirement decreases with increasing $m$ (or, equivalently, with 
increasing $n$) even in the case of complete synchronization. 
Finally, we would like to mention that the 
value of $B_{0,N}$ is non-monotonous with respect to $m$, even 
though it eventually decreases to zero (see Figure~\ref{fig:Bon}). 
Curiously enough, the experiments of \cite{Vu07} with the router, 
running FreeBSD dummynet software, have also shown the 
non-monotonous behavior of the minimal buffer requirement in the 
case of synchronized connections (see Figure~1 in \cite{Vu07}). 
 
\section{Conclusions} 
\label{sec:concl} 
 
In this paper we have studied the interaction between AIMD 
Congestion Control and a bottleneck router with Drop Tail buffer. We 
have used the hybrid modeling approach. It is demonstrated that the 
system always converges to a cyclic behavior. The limit cycles have 
been fully characterized. In particular, we have obtained necessary 
and sufficient condition for the absence of cycles with multiple 
jumps and a simple but tight sufficient condition. Then, we have 
formulated the problem of choosing the buffer size of routers in the 
Internet as a multi-criteria optimization problem. In agreement with 
previous works, our model suggests that as the number of long-lived 
TCP connections sharing the common link increases, the minimal 
buffer size required to achieve full link utilization decreases. 
However, in the case of synchronized connections, the decrease is 
not monotonous and slower than the inverse of the square root of the 
number of connections. The Pareto set obtained with the help of our 
model allows us to evaluate the IP router buffer size in order to 
accommodate real time traffic as well as data traffic. The 
simulations carried out with the help of Simulink and NS Simulator 
confirm the qualitative insights drawn from our model.  
Application of the same framework to other congestion control 
mechanisms, such as MIMD, HighSpeed TCP, TCP Westwood appears to be a 
fruitful direction for future research.

\section*{Appendix} 
 
{\bf Unclipped cycles.} 
 
In this and the next subsection, we ignore the requirement that $y\ge 0$. Thus dynamics is described by equations 
\begin{equation}\label{ea1} 
\left\{\begin{array}{l} 
\frac{dv}{ds}=1; \\ 
\frac{dy}{ds}=\left\{\begin{array}{ll} 
v-y-q, & \mbox{ if } y<b, \mbox{ or } \\ 
 & y=b \mbox{ and } v\le A; \\ 
0 & \mbox{ otherwise,} \end{array} \right. 
\end{array}\right. 
  \end{equation} 
where 
  $$A\defi b+q.$$ 
The jumps occur according to (7) as before. 
 
\begin{defin} Let $y_0=b$ and $v_0<A$ be the initial conditions. A piece of trajectory on the time interval $[0,s^*+1+0]$ is called a {\sl pseudo-cycle} of order $k$ (see (7)). If $v(s^*+1+0)=v_0$ then the pseudo-cycle is called a {\sl $k$-cycle}. 
\end{defin} 
 
Later, it will be shown that if a clipped $k$-cycle exists then the 
unclipped $k$-cycle exists, too (Corollary \ref{co4}). Clearly, 
(\ref{ea1}) has a single solution 
  \begin{equation}\label{ea2} 
\left\{\begin{array}{l} 
v(s)=v_0+s; \\ 
y(s)=(1+q+y_0-v_0)e^{-s}+s-1+v_0-q. 
  \end{array}\right.\end{equation} 
 
\begin{theor} \label{ta1} An (unclipped) $k$-cycle exists iff 
  \begin{equation}\label{ea3} 
A\in\left(\frac{\beta^k}{1-\beta^k},~A^*_k\right], 
  \end{equation} 
where 
  \begin{equation}\label{ea4} 
A^*_k\defi\left\{\begin{array}{ll} 
\frac{\beta^{k-1}(\tau_k+1)}{1-\beta^k}, & \mbox{ if }k>1, \\ 
\infty,& \mbox{ if }k=1 
  \end{array}\right. 
  \end{equation} 
and, for $k>1$, $\tau_k$ is the single positive solution to (\ref{ee6}). 
\end{theor} 
 
\underline{Proof.} Obviously, parameters of a $k$-cycle, $v_0$ and time interval $s_1$ can be found from equations 
  $$y(s_1)=b;~~~~~~~~~~\beta^kv(s_1+1)=v_0,$$ 
which are equivalent to 
\begin{equation}\label{ea5} 
v_0=\frac{\beta^k(s_1+1)}{1-\beta^k}. 
\end{equation} 
  \begin{equation}\label{ea6} 
1-e^{-s_1}=\frac{s_1}{1+A-\frac{\beta^k(s_1+1)}{1-\beta^k}}. 
  \end{equation} 
A $k$-cycle exists iff (\ref{ea6}) has a positive solution and $v_0$ given by (\ref{ea5}) satisfies inequality $v_0\ge \beta A$. (Otherwise, if $v_0<\beta A$, there is no need to reduce $v$ so many times.) Equation (\ref{ea6}) has a positive solution iff 
  $$1+A-\frac{\beta^k}{1-\beta^k}>0 
\quad \mbox{and} \quad 
  \frac{d}{ds}\left.\left[\frac{s}{1+A-\frac{\beta^k(s+1)}{1-\beta^k}}\right]\right|_{s=0}~~<~1$$ 
 
\begin{figure}[h,b] 
\begin{center} 
\begin{picture}(220,220) 
\linethickness{0.5mm} 
\qbezier(0,2)(70,50)(100,200) 
\linethickness{0.1mm} 
\put(10,15){\vector(1,0){180}} 
\put(190,5){$s$} 
\put(15,10){\vector(0,1){190}} 
\put(0,20){$0$} 
\multiput(110,5)(0,10){15}{\line(0,1){5}} 
\linethickness{0.5mm} 
\qbezier(8,0)(50,100)(200,100) 
\linethickness{0.1mm} 
\multiput(10,105)(10,0){15}{\line(1,0){5}} 
\put(0,100){$1$} 
\put(150,80){$1-e^{-s}$} 
\put(100,180){$\frac{s}{1+A-\frac{\beta^k(s+1)}{1-\beta^k}}$} 
\end{picture} 
\caption{Graphical solution to equation (\ref{ea6}).} 
\label{fig0} 
\end{center} 
\end{figure}
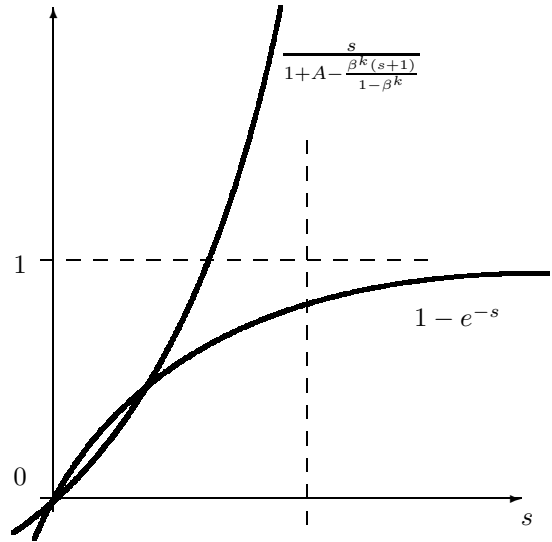 
 
\noindent  
(see Fig.\ref{fig0}), or, equivalently, iff 
  \begin{equation}\label{ea7} 
\beta^k(1+A)<A. 
  \end{equation} 
Put 
  \begin{equation}\label{ea8} 
K\defi\min\{i\ge 1:~~\beta^i<\frac{A}{1+A}\}. 
  \end{equation} 
Before proceeding further, we need the following statements. 
 
\begin{lemma}\label{l1} If $v_0\in[\beta A,A)$ then, starting from $v_0$, $y_0=b$, the next instant series of $K+1$ jumps results in the value $v<A$. Hence the order of any cycle cannot exceed $K+1$ (and clearly cannot be smaller than $K$).\end{lemma} 
 
\underline{Proof.} Suppose $\hat v_0=\beta A$. Then, after the next instant series of $K+1$ jumps, the value $\hat v$ is not smaller than $v$. To put it differently, 
  \begin{equation}\label{ea9} 
v\le \beta^{K+1}[\beta A+\hat s+1], 
  \end{equation} 
where $\hat s$ solves equation 
  $$(1+A-\beta A)e^{-s}+s-1+\beta A=A$$ 
\begin{equation}\label{ea10} 
\Longleftrightarrow\frac{s}{1+(1-\beta)A}-1+e^{-s}=0. 
  \end{equation} 
If we substitute 
  $$\tilde s\defi\frac{A+1}{\beta}-\beta A-1<\frac{A}{\beta^{K+1}}-\beta A-1$$ 
into (\ref{ea10}) we obtain, using equality $A=\frac{\beta\tilde s+\beta-1}{1-\beta^2}$: 
  $$\frac{\tilde s}{1+(1-\beta)A}-1+e^{-\tilde s}=\frac{\tilde s(1+\beta)}{\beta(2+\tilde s)}-1+e^{-\tilde s}>\frac{2\tilde s}{1+\tilde s}-1+e^{-\tilde s}>0.$$ 
When $s$ increases from zero, the lefthand side of (\ref{ea10}) initially decreases from zero and increases thereafter. Hence $\tilde s>\hat s$ and (\ref{ea9}) implies 
  $$v<\beta^{K+1}[\beta A+\tilde s+1]<\beta^{K+1}[\beta A+\frac{A}{\beta^{K+1}}-\beta A-1+1]=A.$$ 
\blacksquare 
 
\begin{lemma}\label{l2} Suppose $\beta\in(0,1)$ is fixed and consider function 
  \begin{equation}\label{ea11} 
f_k(A)\defi (1-\beta^k)A-\beta^{k-1}(s_1+1), 
  \end{equation} 
where $s_1$ solves (\ref{ea6}). The domain of $f$ is given by (\ref{ea7}). Then 
 
(a) $\frac{df_k(A)}{dA}>0$; 
 
(b) $f_1(A)<0$ for all $A>\frac{\beta}{1-\beta}$; 
 
(c) $\forall k>1$ equation $f_k(A)=0$ has a single finite solution $A^*_k$ given by (\ref{ea4}); $A^*_k$ decreases as $k$ increases. 
 
(d)  $\forall k>1$  $A^*_k>\frac{\beta^{k-1}}{1-\beta^{k-1}}$; $\forall k>2$, $A^*_k\le \frac{\beta^{k-2}}{1-\beta^{k-2}}$. 
\end{lemma} 
 
\underline{Proof}. (a) According to the rule of implicit 
differentiation, applied to equation 
  $$\left(1+A-\frac{\beta^k(s_1+1)}{1-\beta^k}\right)\left(1-e^{-s_1}\right)-s_1=0,$$ 
we have 
  $$\frac{ds_1}{dA}=-\frac{(1-e^{-s_1})^2(1-\beta^k)}{e^{-s_1}s_1(1-\beta^k)-(1-e^{-s_1})^2\beta^k-(1-\beta^k)(1-e^{-s_1})}.$$ 
The denominator equals 
  $$-(1-e^{-s_1}-s_1e^{-s_1})-\beta^k(s_1 e^{-s_1}-e^{-s_1}+e^{-2s_1})$$ 
  $$<\beta^k(e^{-s_1}-e^{-2s_1}-s_1e^{-2s_1})-(1-e^{-s_1}-s_1e^{-s_1})$$ 
  $$=(1-e^{-s_1}-s_1e^{-s_1})(\beta e^{-s_1}-1)<0;$$ 
hence $\frac{ds_1}{dA}>0$ for $s_1>0$. 
 
Now 
  $$\frac{df_k(A)}{dA}=(1-\beta^k)-\beta^{k-1}\frac{ds_1}{dA}$$ 
  $$=\frac{(1-\beta^k)[(1-\beta^k)(s_1e^{-s_1}-1+e^{-s_1})+\beta^{k-1}(1-e^{-s_1})^2(1-\beta)]} 
{s_1e^{-s_1}(1-\beta^k)+(1-e^{-s_1})(\beta^k e^{-s_1}-1)}.$$ 
The denominator is negative (see above). The nominator does not exceed 
  $$(1-\beta^k)(1-\beta)[(s_1 e^{-s_1}-1+e^{-s_1})(1+\beta^{k-1})+\beta^{k-1}(1-e^{-s_1})^2]$$ 
  $$=(1-\beta^k)(1-\beta)[(s_1 e^{-s_1}-1+e^{-s_1})+\beta^{k-1}e^{-s_1}(s_1-1+e^{-s_1})].$$ 
The both terms in the latter square bracket are negative for $s_1>0$. Hence $\frac{df_k(A)}{dA}>0$. 
 
(b) It is sufficient to prove that 
   $$s_1>S\defi A(1-\beta)-1,$$ 
where $s_1$ solves (\ref{ea6}) at $k=1$. 
 
Case $S<0$ is trivial, thus assume that $S>0$. Let us substitute $S$ into the both sides of (\ref{ea6}) and estimate the difference: 
  $$\frac{S}{1+A-\frac{\beta(S+1)}{1-\beta}}-1+e^{-S}=\frac{S}{1+\frac{S+1}{1-\beta}-\frac{\beta(S+1)}{1-\beta}}-1+e^{-S} 
  =e^{-S}-\frac{2}{S+2}<0,$$ 
because function $(S+2)e^{-S}$ decreases from 2 at $S=0$. To complete this part  of the proof, it is sufficient to notice that, on the interval 
  $$0<S<\frac{A(1-\beta)}{\beta}+\frac{1-2\beta}{\beta},$$ 
the righthand side of (\ref{ea6}) is smaller than the lefthandside iff $S<s_1$. 
 
(c) The first part is obvious: $A^*_k$ is given by (\ref{ea4}), provided equation (\ref{ee6}) has a single positive solution. The latter statement folows from the fact that function 
  $$g(\tau)=(1-e^{-\tau})(1+\alpha(\tau+1))/\tau$$ 
decreases to $\lim_{\tau\to\infty} g(\tau)=\alpha$, starting from $\lim_{\tau\to 0}g(\tau)=1+\alpha$. Here 
  \begin{equation}\label{ea12} 
\alpha\defi\frac{\beta^{k-1}-\beta^k}{1-\beta^k}. 
  \end{equation} 
Indeed, 
  $$\frac{dg}{d\tau}=\frac{e^{-\tau}[1+\alpha+\tau(1+\alpha+\alpha\tau)]-(1+\alpha)}{\tau^2}<0$$ 
in case $\alpha<1$, and 
  \begin{equation}\label{ea13} 
\alpha=\frac{\beta^{k-1}}{1+\beta+\ldots+\beta^{k-1}}\le\frac{1}{1+1/\beta}<1/2. 
  \end{equation} 
 
Now, look what happens as $k$ increases. Obviously, functions 
$\frac{\beta^{k-1}}{1-\beta^k}=\frac{\beta^k}{1-\beta^k}\cdot\frac{1}{\beta}$ 
and $\alpha=\frac{\beta^k}{1-\beta^k}(\frac{1}{\beta}-1)$ (see 
(\ref{ea12})) decrease. According to (\ref{ea4}) it remains to prove 
that $\tau_k$ given by (\ref{ee6}) increases with $\alpha$. We 
rewrite (\ref{ee6}) as $(1+\alpha(\tau+1))(1-e^{-\tau})-\tau=0.$ 
Hence 
  $$\frac{d\tau_k}{d\alpha}=-\frac{(\tau_k+1)(1-e^{-\tau_k})}{\alpha(1-e^{-\tau_k})+(1+\alpha(\tau_k+1))e^{-\tau_k}-1}=-\frac{(\tau_k+1)^2(1-e^{-\tau_k})^2}{h(\tau_k)},$$ 
where 
  $h(\tau)=-2+3e^{-\tau}-e^{-2\tau}+\tau e^{-\tau}+\tau^2 e^{-\tau}.$ 
(We have substituted $\alpha=\frac{\tau_k-1+e^{-\tau_k}}{(1-e^{-\tau_k})(\tau_k+1)}$.) We intend to prove that 
  \begin{equation}\label{ea14} 
\frac{dh}{d\tau}=-2e^{-\tau}+2 e^{-2\tau}+\tau e^{-\tau}-\tau^2 e^{-\tau}<0 
  \end{equation} 
when $\tau>0$. Clearly (\ref{ea14}) holds for $\tau\ge 1$. 
 
Suppose $\tau\in(0,1)$. Then 
  $$\frac{d^2 h}{d\tau^2}=3e^{-\tau}-4 e^{-2\tau}-3\tau e^{-\tau}+\tau^2 e^{-\tau} 
  <3e^{-\tau}-4 e^{-2\tau}-2\tau e^{-\tau}=e^{-\tau}(3-4 e^{-\tau}-2\tau).$$ 
Expression in the brackets has a negative maximum at $\tau=\ln 2$. Therefore, $\frac{d^2 h}{d\tau^2}<0$ and $\frac{dh}{d\tau}<0$. Finally, $h(\tau)<0$ for all $\tau>0$, because $h(0)=0$. 
 
(d) To estimate $A^*_k$ from below, we use statement (a): it is sufficient to establish that $f_k\left(\frac{\beta^{k-1}}{1-\beta^{k-1}}\right)<0$, ie $s_1+1>\frac{1-\beta^k}{1-\beta^{k-1}}\Longleftrightarrow \frac{S}{1+\frac{\beta^{k-1}}{1-\beta^{k-1}}-\frac{\beta^k(S+1)}{1-\beta^k}}<1-e^{-S}$ for $S=\frac{1-\beta^k}{1-\beta^{k-1}}-1$. (The argument is similar to (b).) But 
  $$\frac{S}{\frac{1}{1-\beta^{k-1}}-\frac{\beta^k}{1-\beta^k}\cdot\frac{1-\beta^k}{1-\beta^{k-1}}}-1+e^{-S}=\frac{S}{S+1}-1+e^{-S}=e^{-S}-\frac{1}{1+S}<0$$ 
because function $e^{-S}(1+S)$ decreases from $1$ at $S=0$. 
 
Finally, in case $k>2$, suppose $A^*_k>\frac{\beta^{k-2}}{1-\beta^{k-2}}$. Then for parameters values $\beta$ and $A\in\left(\frac{\beta^{k-2}}{1-\beta^{k-2}}, A^*_k\right)$ we have that (\ref{ea7}) holds for $k-2$, $k-1$, and $k$ and simultaneously $f_k(A)<0$, $f_{k-1}(A)<0$, $f_{k-2}(A)<0$: see (a) and (c). According to the beginning of the proof of Theorem 1, cycles of orders $k$, $k-1$, and $k-2$ exist which contradicts Lemma \ref {l1}. \blacksquare 
 
Now we can easily finish the \underline{proof of Theorem \ref{ta1}}. Suppose a $k$-cycle exists. Then, according to (\ref{ea7}), $A>\frac{\beta^k}{1-\beta^k}$. Lemma \ref{l2} guarantees that 
  $$A^*_k>\frac{\beta^{k-1}}{1-\beta^{k-1}}>\frac{\beta^k}{1-\beta^k},$$ 
and, as was mentioned earlier, inequality $v_0\ge \beta A$ must be valid (see (\ref{ea5})), which is equivalent to $A\le A^*_k$. Finally, if (\ref{ea3}) holds then (\ref{ea6}) has a positive solution (see (\ref{ea7}) ) and $v_0\ge \beta A$; hence a $k$-cycle exists. \blacksquare 
\begin{figure}[h] 
\begin{center} 
\begin{picture}(350,200) 
\put(12,15){\vector(1,0){300}} 
\put(15,12){\vector(0,1){170}} 
\put(0,190){\footnotesize\it Case $A^*_{N+1}\ge q$:} 
\put(2,188){\vector(1,-1){10}} 
\multiput(80,12)(0,10){16}{\line(0,1){5}} 
\put(80,170){\vector(0,1){15}} 
\put(70,190){\footnotesize\it Case $A^*_{N+1}<q$:} 
\put(93,188){\vector(-1,-1){10}} 
\put(0,0){$0$} 
\put(20,175){\footnotesize 5-cycle exists} 
\linethickness{0.5mm} 
\put(0,170){\line(1,0){60}} 
\put(60,169.5){\circle*{3}} 
\linethickness{0.1mm} 
\multiput(60,170)(0,-10){16}{\line(0,-1){5}} 
\put(45,0){\protect\footnotesize$A^*_5-q$} 
\linethickness{0.5mm} 
\put(5,140){\line(1,0){120}} 
\put(125,139.5){\circle*{3}} 
\put(30,130){\footnotesize 4-cycle} 
\put(30,120){\footnotesize exists} 
\linethickness{0.1mm} 
\multiput(125,140)(0,-10){13}{\line(0,-1){5}} 
\put(110,0){\protect\footnotesize$A^*_4-q$} 
\linethickness{0.5mm} 
\put(117,108){\footnotesize 3-cycle exists} 
\put(200,100){\vector(-1,0){100}} 
\put(200,99.5){\circle*{3}} 
\linethickness{0.1mm} 
\multiput(200,100)(0,-10){9}{\line(0,-1){5}} 
\put(185,0){\protect\footnotesize$A^*_3-q$} 
\multiput(100,100)(0,-10){6}{\line(0,-1){5}} 
\put(85,25){\footnotesize$\frac{\beta^3}{1-\beta^3}-q$} 
\linethickness{0.5mm} 
\put(180,78){\footnotesize 2-cycle exists} 
\put(270,70){\vector(-1,0){100}} 
\put(270,69.5){\circle*{3}} 
\linethickness{0.1mm} 
\multiput(270,70)(0,-10){6}{\line(0,-1){5}} 
\put(255,0){\footnotesize$A^*_2-q$} 
\put(275,30){\footnotesize 1-cycle exists} 
\multiput(170,70)(0,-10){3}{\line(0,-1){5}} 
\put(155,25){\footnotesize$\frac{\beta^2}{1-\beta^2}-q$} 
\linethickness{0.5mm} 
\put(340,40){\vector(-1,0){100}} 
\put(346,40){\circle*{2}} 
\put(350,40){\circle*{2}} 
\linethickness{0.1mm} 
\put(240,40){\line(0,-1){5}} 
\put(225,25){\footnotesize$\frac{\beta}{1-\beta}-q$} 
\put(312,0){$b$} 
\end{picture} 
\caption{Existence of unclipped cycles; $N=4$.} 
\label{fig3} 
\end{center} 
\end{figure} 
 
Remember that $A=b+q$. Thus, if $q$ is fixed and $b$ increases from $0$, unclipped cycles have orders $N$ (see (\ref{ee1})) and,  possibly, $N+1$, if $A^*_{N+1}-q>0$.  Later, as $b$ increases, the order of cycles decreases according to Fig. \ref{fig3}.

{\bf Stability of unclipped cycles.} 
 
We intend to study the mapping $\varphi$ introduced just before 
Theorem~\ref{prop_stab}. Since we study only unclipped cycles, this 
map is a little different and will be denoted $\tilde\varphi$. But 
firstly we concentrate on a different mapping: 
  $$\Phi^k(v_0)=\beta^k(v_0+s^*+1)$$ 
defined for $v_0\in[\beta A,A]$ under a fixed $k\ge 1$. Here $s^*\defi 0$ if $v_0=A$; in case $v_0<A$, $s^*>0$ is the first moment when $y(s^*)=b$ starting from $y(0)=b$, $v(0)=v_0$. 
 
\begin{lemma}\label{l3} $\left|\frac{d\Phi^k(v_0)}{dv_0}\right|<\beta^k$ and hence $\Phi^k$ is a contraction. Function $\Phi^k$ is decreasing. \end{lemma} 
 
\underline{Proof.} Assuming that $v_0<A$, $s^*$ is a single positive solution to equation 
  \begin{equation}\label{ea15} 
(1+A-v_0)(1-e^{-s})-s=0, 
  \end{equation} 
hence 
  $$\frac{ds^*}{dv_0}=\frac{1-e^{-s^*}}{(1+A-v_0)e^{-s^*}-1}=\frac{1-e^{-s^*}}{\frac{s^*}{1-e^{-s^*}}\cdot e^{-s^*}-1}$$ 
and 
  $$\frac{d\Phi^k}{dv_0}=\beta^k(1+\frac{ds^*}{dv_0})=\beta^k e^{-s^*}\frac{s^*-1+e^{-s^*}}{s^*e^{-s^*}+e^{-s^*}-1}<0.$$ 
Finally, 
  $$e^{-s^*}\frac{s^*-1+e^{-s^*}}{s^*e^{-s^*}+e^{-s^*}-1}+1=e^{-s^*}\frac{e^{s^*}-e^{-s^*}-2s^*}{1-e^{-s^*}-s^*e^{-s^*}}>0,$$ 
because the nominator increases, starting from $0$ at $s^*=0$. Therefore $\frac{d\Phi^k}{dv_0}>(-\beta^k)$. \blacksquare 
 
\begin{figure}[t,h] 
\begin{center} 
\begin{picture}(420,220) 
\put(20,210){(a)} 
\linethickness{0.5mm} 
\qbezier(25,125)(100,60)(170,50) 
\linethickness{0.1mm} 
\put(10,15){\vector(1,0){180}} 
\put(15,10){\vector(0,1){190}} 
\multiput(25,15)(0,10){11}{\line(0,1){5}} 
\multiput(15,25)(10,0){10}{\line(1,0){5}} 
\multiput(170,15)(0,10){16}{\line(0,1){5}} 
\multiput(15,170)(10,0){16}{\line(1,0){5}} 
\put(25,125){\circle*{2}} 
\put(160,50){\line(1,0){10}} 
\put(160,55){\line(2,-1){10}} 
\put(10,10){\line(1,1){170}} 
\multiput(83,80)(0,-10){7}{\line(0,-1){5}} 
\put(3,3){$0$} 
\put(-3,20){$\beta A$} 
\put(0,165){$A$} 
\put(5,190){z} 
\put(20,0){$\beta A$} 
\put(165,0){$A$} 
\put(190,5){$v_0$} 
\put(80,0){$V_2$} 
\put(140,180){$z=v_0$} 
\put(105,85){$z=\tilde\varphi(v_0)$} 
\put(115,70){$=\Phi^K(v_0)$} 
\put(240,210){(b)} 
\put(230,15){\vector(1,0){180}} 
\put(235,10){\vector(0,1){190}} 
\multiput(245,15)(0,10){9}{\line(0,1){5}} 
\multiput(235,25)(10,0){10}{\line(1,0){5}} 
\multiput(390,15)(0,10){16}{\line(0,1){5}} 
\multiput(235,170)(10,0){16}{\line(1,0){5}} 
\multiput(310,15)(0,10){16}{\line(0,1){5}} 
\multiput(235,90)(10,0){16}{\line(1,0){5}} 
\put(230,10){\line(1,1){170}} 
\linethickness{0.5mm} 
\qbezier(310,170)(340,130)(390,110) 
\linethickness{0.1mm} 
\put(310,170){\line(2,-1){10}} 
\put(310,170){\line(1,-3){5}} 
\put(380,110){\line(1,0){10}} 
\put(380,120){\line(1,-1){10}} 
\linethickness{0.5mm} 
\qbezier(245,65)(280,30)(310,25) 
\linethickness{0.1mm} 
\put(245,65){\circle*{2}} 
\put(310,25){\circle*{2}} 
\put(223,3){$0$} 
\put(217,20){$\beta A$} 
\put(220,165){$A$} 
\put(225,190){z} 
\put(235,0){$\beta A$} 
\put(385,0){$A$} 
\put(410,5){$v_0$} 
\put(305,0){$d$} 
\multiput(267,20)(0,10){3}{\line(0,1){5}} 
\multiput(350,15)(0,10){12}{\line(0,1){5}} 
\put(220,85){$d$} 
\put(265,0){$V_1$} 
\put(345,0){$V_2$} 
\put(360,180){$z=v_0$} 
\put(242,127){$z=\tilde\varphi(v_0)=\Phi^K(v_0)$} 
\put(277,43){$z=\tilde\varphi(v_0)=\Phi^{K+1}(v_0)$} 
\end{picture} 
\caption{Graphs of $\tilde\varphi(v_0)$.} 
\label{fig4} 
\end{center} 
\end{figure}
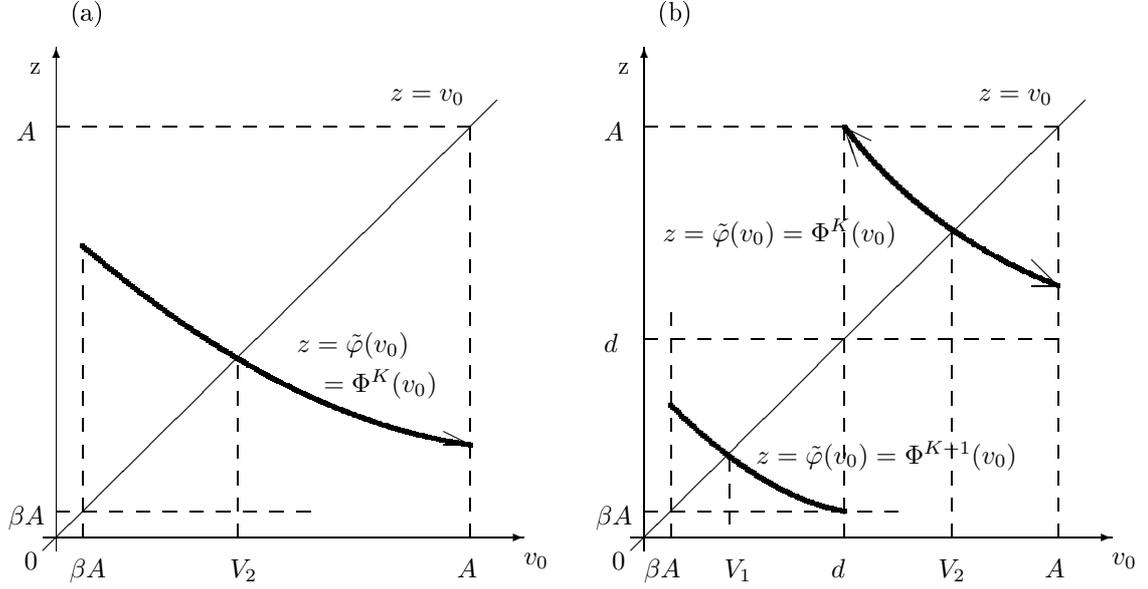

\begin{lemma}\label{l4} (a) $A\in\left(A^*_{K+1},\frac{\beta^{K-1}}{1-\beta^{K-1}}\right]$ iff $d<\beta A$, where $d$ is a solution to $\Phi^K(d)=A$. Here and below, $\frac{\beta^{K-1}}{1-\beta^{K-1}}\defi\infty$ if $K=1$; $K$ is defined by (\ref{ea8}). 
 
In this case, $\forall v_0\in[\beta A,A)$, the mapping $\tilde\varphi(v_0)$ coincides with $\Phi^K(v_0)$. 
 
(b) If $A\in\left(\frac{\beta^K}{1-\beta^K},A^*_{K+1}\right]$ the following statements hold: 
 
\ \ \ \ \ ($\alpha$) $\forall v_0\in[\beta A,d]$, $\tilde\varphi(v_0)=\Phi^{K+1}(v_0)\in[\beta A,d]$; 
 
\ \ \ \ \ ($\beta$) $\forall v_0\in (d,A)$, $\tilde\varphi(v_0)=\Phi^K(v_0)\in(d,A)$. 
 
See Fig.\ref{fig4}. (Note that, according to the definition of $K$, $A\in\left(\frac{\beta^K}{1-\beta^K},\frac{\beta^{K-1}}{1-\beta^{K-1}}\right]$; according to Lemma \ref{l2}, $A^*_{K+1}\in\left(\frac{\beta^K}{1-\beta^K},\frac{\beta^{K-1}}{1-\beta^{K-1}}\right]$.) \end{lemma} 
 
\underline{Proof.} (a) According to the definition, $d=\frac{A}{\beta^K}-s^*-1$, where $s^*$ solves (\ref{ea15}) under $v_0=d$. If $d=\beta A$ then 
  $$\left\{\begin{array}{rcl} 
(1+A-\beta A)(1-e^{-s^*}) & = & s^*;\\ 
\frac{A}{\beta^K}-s^*-1 & = & \beta A, 
  \end{array}\right.$$ 
or equivalently 
  $$\left\{\begin{array}{rcl} 
A & = & \frac{\beta^K(s^*+1)}{1-\beta^{K+1}};\\ 
(1+A-\frac{\beta^{K+1}(s^*+1)}{1-\beta^{K+1}})(1-e^{-s^*}) & = & s^*. 
  \end{array}\right.$$ 
To put it differently, we have $A=A^*_{K+1}$ if $d=\beta A$. 
 
It remains to prove that $d-\beta A=\frac{A}{\beta^K}-s^*-1-\beta A$ is a decreasing function of $A$. Since $s^*$ satisfies equation 
  $$\left(1+A-\frac{A}{\beta^K}+s^*+1\right)(1-e^{-s^*})-s^*=0,$$ 
  $$\frac{ds^*}{dA}=\frac{(1-\beta^K)(1-e^{-s^*})^2}{\beta^Ke^{-s^*}(e^{-s^*}+s^*-1)}$$ 
and 
  $$\frac{d(d-\beta A)}{dA}=\frac{1}{\beta^K}-\frac{ds^*}{dA}-\beta 
  =\frac{s^*e^{-s^*}+e^{-s^*}-1+\beta^K(1-e^{-s^*})^2-\beta^{K+1}e^{-s^*}(e^{-s^*}+s^*-1)}{\beta^Ke^{-s^*}(e^{-s^*}+s^*-1)}.$$ 
The denominator is obviously positive for $s^*>0$. The nominator equals zero when $s^*=0$, its derivative equals 
  $$e^{-s^*}[-s^*+2\beta^K(1-e^{-s^*})-\beta^{K+1}(2-2e^{-s^*}-s^*)].$$ 
Expression in the square brackets equals zero when $s^*=0$ and has derivative 
  $$-1+2\beta^Ke^{-s^*}-2\beta^{K+1}e^{-s^*}+\beta^{K+1}\defi g(s^*,\beta).$$ 
Clearly, 
  $$\frac{\partial g(s^*,\beta)}{\partial s^*}=2\beta^K e^{-s^*}(\beta-1)<0,$$ 
and finally $g(0,\beta)=-1+2\beta^K-\beta^{K+1}<0$ for all $\beta\in(0,1)$ because $g(0,1)=0$ and $\frac{dg(0,\beta)}{d\beta}=\beta^{K-1}[K(1-\beta)+K-\beta]>0$. Therefore $\frac{d(d-\beta A)}{dA}<0$. 
 
According to Lemma \ref{l1}, $\tilde\varphi$ can coincide with $\Phi^K$ or $\Phi^{K+1}$ only. In case (a), $\Phi^K(A)<A$ because $\lim_{v_0\to A} s^*=0$ (see (\ref{ea8}) ). Function $\Phi^K$ increases as $v_0$ decreases (Lemma \ref{l3}), but $\Phi^K(v_0)=A$ when $v_0=d<\beta A$.  Thus, $\forall v_0\in[\beta A,A)$ $\Phi^K(v_0)<A$, $(K+1)$ instant jumps are never needed and $\tilde\varphi=\Phi^K$. 
 
(b) In this case, $d\ge\beta A$ according to (a). Since $\Phi^K(d)=A$ and $\Phi^K$ is a decreasing function (Lemma \ref{l3}), $\Phi^K(v_0)\ge A$ if $v_0\in[\beta A,d]$ and $\tilde\varphi(v_0)=\Phi^{K+1}(v_0)$, as $K$ jumps are not sufficient. Obviously, $\tilde\varphi(d)=\Phi^{K+1}(d)=\beta A$. Now 
  $$\tilde\varphi(\beta A)=\Phi^{K+1}(\beta A)=\Phi^{K+1}(d)-\int_{\beta A}^d \frac{d\Phi^{K+1}(v_0)}{dv_0} dv_0 
  <\Phi^{K+1}(d)+(d-\beta A)=d$$ 
according to Lemma \ref{l3}, and statement ($\alpha$) is proved. 
 
In case ($\beta$), $\Phi^K(v_0)<A$, hence $\tilde\varphi(v_0)=\Phi^K(v_0)$. We know that $\Phi^K(d)=A$. Using Lemma \ref{l3}, we conclude that 
  $$\Phi^K(A)=\Phi^K(d)+\int_d^A \frac{d\Phi^K(v_0)}{dv_0} dv_0>A-(A-d)=d.$$ 
\blacksquare 
 
\begin{conse}\label{co2} Theorem~\ref{prop_stab} and Corollary \ref{cor_simple_cycle} hold for unclipped cycles. \end{conse} 
 
\underline{Proof.} (See Fig.\ref{fig4}.) Under conditions (a) of Lemma \ref{l4}, $\tilde\varphi$ has a stable stationary point $V_2$ coincident with that of $\Phi^K$.  (Note that $\Phi^K(A)<A$, so that $V_2\in[\beta A,A)$.) 
 
Consider case (b) of Lemma \ref{l4}. 
 
If $v_0\in[\beta A,d]$ then $\tilde\varphi=\Phi^{K+1}$ is a contraction defined on this interval; so that the statement follows. 
 
If $v_0\in(d,A)$, $\tilde\varphi$ has a stable stationary point $V_2$ coincident with that of $\Phi^K$. (Note that $\Phi^K(A)<A$, hence $d<A=\Phi^K(d)$, so that $V_2\in(d,A)$.) 
 
Corollary \ref{cor_simple_cycle} is obvious. \blacksquare 
 
{\bf Critical cycles.} 
 
Remind that a cycle is called critical if $\min_s y(s)=0$ From (\ref{ea2},\ref{ea5},\ref{ea6}) it is clear that the minimum is attained at 
  \begin{equation}\label{ea16} 
s_0=\ln\frac{s_1}{1-e^{-s_1}}, 
  \end{equation} 
where $s_1$ solves (\ref{ea6}). 
 
\begin{lemma}\label{l5} Suppose, an unclipped $k$-cycle exists. 
 
(a) $y(s_0)$ increases with $A$. 
 
(b) For cycles of order $k=1$, $\exists\varepsilon>0$ $\exists\delta>0$: $\frac{dy(s_0)}{dA}>\varepsilon$ as soon as $A>\frac{\beta}{1-\beta}+\delta$. Consequently $y(s_0)\to\infty$ as $A\to\infty$. 
\end{lemma} 
 
\underline{Proof.} (a) After rewriting (\ref{ea6}) in the form 
  $$\left(1+A-\frac{\beta^k(s_1+1)}{1-\beta^k}\right)(1-e^{-s_1})-s_1=0,$$ 
we obtain: 
\begin{equation}\label{eaadd} 
  \frac{ds_1}{dA}=\frac{1-e^{-s_1}}{\frac{\beta^k}{1-\beta^k}(1-e^{-s_1})-e^{-s_1}\frac{s_1}{1-e^{-s_1}}+1} 
=\frac{(1-e^{-s_1})^2(1-\beta^k)}{1-e^{-s_1}-s_1 e^{-s_1}(1-\beta^k)-\beta^k e^{-s_1}+\beta^ke^{-2s_1}}. 
\end{equation} 
The denominator has derivative (wrt $s_1>0$) 
  $$s_1e^{-s_1}(1-\beta^k)+2\beta^k(e^{-s_1}-e^{-2s_1})>0$$ 
and hence increases starting from $0$ when $s_1=0$. Therefore $\frac{ds_1}{dA}>0$. 
 
Since 
  \begin{equation}\label{ea17} 
y(s_0)=(1+A-v_0)e^{-s_0}+s_0-1+v_0-q=v_0+s_0-q 
  \end{equation} 
we conclude that 
  $$\frac{dy(s_0)}{dA}=\left(\frac{dv_0}{ds_1}+\frac{ds_0}{d s_1}\right)\frac{ds_1}{dA} 
  =\left(\frac{\beta^k}{1-\beta^k}+\frac{1-e^{-s_1}-s_1e^{-s_1}}{s_1(1-e^{-s_1})}\right)\frac{ds_1}{dA}>0.$$ 
 
(b) Note that the denominator in (\ref{eaadd}) is a bounded function of $s_1$. Thus $\exists\varepsilon>0$ $\exists \delta_1>0$: $\frac{ds_1}{dA}>\varepsilon$ as soon as $s_1>\delta_1$, or, equivalently, as soon as $A>\frac{\beta}{1-\beta}+\delta$, where $\delta>0$ exists because $s_1$ monotonically increases with $A$. 
Remember that $\lim_{A\to\frac{\beta}{1-\beta}} s_1=0$. 
\blacksquare 
 
\begin{lemma}\label{l6} Suppose, all parameters, apart from $b$, are fixed. 
 
(a) A critical cycle of order $k$ exists (for some positive value of $b$) if and only if 
\begin{equation}\label{ea18} 
\frac{\beta^k}{1-\beta^k}<q\le q^*_k, 
\end{equation} 
where $q^*_k$ is given by (\ref{ee8}). 
The corresponding value of $b$ equals $b_{0,k}$, see (\ref{ee5}). 
 
(b) The boundary $q^*_k$ satisfies inequalities 
  \begin{equation}\label{e21} 
\frac{\beta^{k-1}}{1-\beta^{k-1}}\le q^*_k<A^*_k. 
  \end{equation} 
(In case $k=1$, $q^*_1=+\infty$.) 
\end{lemma} 
 
\underline{Proof.} (a) Necessity. Let $k>1$ and suppose a critical cycle of order $k$ exists. Then, if we increase $b$ up to $b^*=A^*_k-q$, this $k$-cycle (equipped with an asterisk) must remain unclipped (Lemma \ref{l5}): 
  \begin{equation}\label{ea19} 
y^*(s_0^*)=v^*_0+s^*_0-q\ge 0 
  \end{equation} 
(see (\ref{ea17}) ), ie $q\le v^*_0+s^*_0$. Here $s^*_0=\ln\frac{s^*_1}{1-e^{-s^*_1}}$ (see \ref{ea16}) ), $s^*_1$ solves (\ref{ea6}) under $A^*_k$ and hence coincides with $\tau_k$ defined by (\ref{ee6}); $v^*_0$ is defied by (\ref{ea5}). Therefore, $v^*_0+s^*_0=q^*_k$. 
 
Obviously, system of equations (\ref{ea5},\ref{ea6},\ref{ea16}) and 
  $$v_0+s_0-q=0$$ 
(see (\ref{ea17}) ) must be compatible, ie equation 
  \begin{equation}\label{ea20} 
h(s_1)=\frac{\beta^k(s_1+1)}{1-\beta^k}+\ln\frac{s_1}{1-e^{-s_1}}-q=0 
  \end{equation} 
must have a positive solution. One can easily check that $h$ increases to infinity with $s_1$, starting from $\lim_{s_1\to 0}h(s_1)=\frac{\beta^k}{1-\beta^k}-q$. Hence $q>\frac{\beta^k}{1-\beta^k}$. 
 
In case $k=1$ we put $q^*_1=+\infty$, so that (\ref{ea18}) transforms to $q>\frac{\beta}{1-\beta}$, and the proof of the latter inequality remains unchanged. 
 
Before proving sufficiency, we firstly prove part (b). 
 
(b) Let $k>1$; 
  $$h\defi q^*_k-\frac{\beta^{k-1}}{1-\beta^{k-1}}=\frac{\beta^k}{1-\beta^k}(\tau_k+1)+\ln\frac{\tau_k}{1-e^{-\tau_k}}-\frac{\beta^{k-1}}{1-\beta^{k-1}}$$ 
  $$=\ln\frac{\tau_k}{1-e^{-\tau_k}}-\alpha(\tau_k+1)-\alpha(\tau_k+1)\gamma+\tau_k\gamma,$$ 
where $\alpha\defi\frac{\beta^{k-1}-\beta^k}{1-\beta^k}$, $\gamma\defi\frac{\beta^{k-1}}{1-\beta^{k-1}}$. Using (\ref{ee6}), the last expression can be rewritten as 
  $$h=1-\frac{\tau_k}{1-e^{-\tau_k}}+\ln\frac{\tau_k}{1-e^{-\tau_k}}+\left(1-\frac{\tau_k}{1-e^{-\tau_k}}\right)\gamma+\tau_k\gamma.$$ 
For $k>1$ one can easily check that $\gamma\ge\frac{\alpha}{1-2\alpha}$; therefore, since $1-e^{-\tau_k}-\tau_ke^{-\tau_k}\ge 0$, 
  \begin{equation}\label{ea21}\begin{array}{c}\displaystyle 
h\ge 1-\frac{\tau_k}{1-e^{-\tau_k}}+\ln\frac{\tau_k}{1-e^{-\tau_k}}+\frac{1-e^{-\tau_k}-\tau_ke^{-\tau_k}}{1-e^{-\tau_k}}\cdot\frac{\alpha}{1-2\alpha}\\ 
\  \\ 
\displaystyle =1-\frac{\tau_k}{1-e^{-\tau_k}}+\ln\frac{\tau_k}{1-e^{-\tau_k}}+\frac{1-e^{-\tau_k}-\tau_ke^{-\tau_k}}{1-e^{-\tau_k}}\cdot 
\frac{\tau_k-1+e^{-\tau_k}}{3-3e^{-\tau_k}-\tau_k-\tau_ke^{-\tau_k}}. 
  \end{array}\end{equation} 
(We have used (\ref{ee6}) to express $\alpha$ in terms of $\tau_k$.) 
 
During the proof of Lemma \ref{l2}(c), we established that $\tau_k$ increases with $\alpha\in(0,1/2)$, starting from $0$ when $\alpha=0$. Hence $\tau_k\in(0,\tau)$, where $\tau$ is the single positive solution to equation 
  $$(1-e^{-\tau})(1+\frac{1}{2}(\tau+1))=\tau.$$ 
(The solvability was established in the Proof of Lemma \ref{l2}(c).) 
 
Now the righthand side of (\ref{ea21}) is non-negative if $\tau_k\in(0,\tau)$. This statement was accuratly checked numerically; the analytical proof is problematic. 
 
The second inequality, to be verified, is obvious: 
  $$q^*_k-A^*_k=\frac{\beta^k}{1-\beta^k}(\tau_k+1)+\ln\frac{\tau_k}{1-e^{-\tau_k}}-\frac{\beta^{k-1}(\tau_k+1)}{1-\beta^k} 
  =\ln[1+\alpha(\tau_k+1)]-\alpha(\tau_k+1)<0.$$ 
 
(a) Sufficiency. Suppose inequalities (\ref{ea18}) hold. Then for $b\in[0,A^*_k-q]$ (unclipped) $k$-cycles exist according to Theorem \ref{ta1}, see Fig.\ref{fig3}. (Remember that $A^*_1=q^*_1=+\infty$.) Note that, in case $k>1$, $q<A^*_k$ due to (b). In this case, for $b=b^*=A^*_k-q$, 
  $$y^*(s^*_0)=v^*_0+s^*_0-q=q^*_k-q\ge 0$$ 
(see (\ref{ea19}) ) and this particular cycle  is really unclipped. In case $k=1$, according to Lemma \ref{l5}(b), $y(s_0)>0$ for sufficiently large $b$. Now, if $b$ decreases then the minimal value of $y$ over a cycle decreases (Lemma \ref{l5}(a)~) and, being continuous, becomes zero, since $y(s_0)<0$ for the unclipped $k$-cycle corrresponding to $b=0$. 
 
To calculate the critical value of $b$, note that equation (\ref{ea20}) has a single positive solution $s_1$. Now, if we take 
  $$b=\frac{s_1}{1-e^{-s_1}}+\frac{\beta^k(s_1+1)}{1-\beta^k}-1-q=\frac{s_1}{1-e^{-s_1}}-\ln\frac{s_1}{1-e^{-s_1}}-1$$ 
then, according to (\ref{ea5},\ref{ea6}), the corrresponding cycle will be critical. (One can easily see that $b>0$.) It remains to notice that equation (\ref{ea20}) is identical with (\ref{ee4}). \blacksquare 
 
\begin{conse}\label{co3} Let $N$ be defined by (\ref{ee1}). Then critical cycles of orders $k<N$ cannot exist. \end{conse} 
 
\underline{Proof.} According to (\ref{ee1}), $q\le\frac{\beta^k}{1-\beta^k}$, if $n<N$. The statement follows from Lemma \ref{l6}(a).  \blacksquare 
 
{\bf Clipped cycles.} 
 
\underline{Proof of Theorem~\ref{prop_stab}.} Let $S$ be the single 
positive solution to equation 
  $$(1+b+S)e^{-S}=1.$$ 
Then a continuous trajectory (\ref{ea2}) starting from $(y_0=b,~v_0=q-S)$ touches the axis $y=0$ at a single point, at time moment $S$. 
 
(a) In case $S>q-\beta A$ it is obvious that starting from any point $(y_0=b,v_0\in[\beta A,A))$, the trajectory never touches the axis $y=0$. The statements follow now from Corollary \ref{co2}: the mappings $\varphi$ and $\tilde\varphi$ coincide. 
 
(b) Suppose that $S\le q-\beta A$ and $q-S<V$, where $V$ ($=V_1$ or $V_2$) is the minimal stationary point of the mapping $\tilde\varphi$ (see Lemma \ref{l4} and Fig.\ref{fig4}). Then, starting from any point $(y_0=b,v_0\in[\beta A,A))$, at most $\varphi(\varphi(v_0))$ is such that the further trajectory never touches the axis $y=0$: see Lemmas \ref{l3} and \ref{l4}. To put it differently, $\varphi^n(v_0)>q-S$ for $n\ge 2$. The required statements again follow from Corollary \ref{co2}. The mappings $\varphi$ and $\tilde\varphi$ coincide on the domain $[q-S,A)$. 
 
(c) Suppose that $S\le q-\beta A$, $q-S\ge V_2$, where $V_2$ is the 
maximal stationary point of the mapping $\tilde\varphi$ (see Lemma 
\ref{l4} and Fig.\ref{fig4}). Then, starting from any point 
$(y_0=b,v_0\in[\beta A,A))$ $\forall n\ge 2,~~ 
\varphi^n(v_0)=\varphi(\varphi(v_0))$ because 
$\varphi(v_0),~\varphi(\varphi(v_0))\le V_2\le q-S$. Note that in 
terms of Theorem~\ref{prop_stab}, $d<\beta(q+b)$, 
$V_2=\varphi(\varphi(v_0))$ is different from (smaller than) $V_2$ 
shown on Fig.\ref{fig4}. 
 
(d) Suppose that $S\le q-\beta A$, case (b) (Lemma \ref{l4}) takes place and $V_1\le q-S\le d$ (see Fig.\ref{fig4}). Then, if $v_0\in[\beta A,d]$, the situation is similar to (c): $\forall n\ge 2$ $\varphi^n(v_0)=\varphi(\varphi(v_0))$, because $\varphi(v_0),~\varphi(\varphi(v_0))\le V_1\le q-S\le d$. 
 
If $v_0\in(d,A)$ then the trajectory never touches axis $y=0$ because $\forall n$ $\varphi^n(v_0)=\tilde\varphi^n(v_0)>d\ge q-S$. The statements follow from Corollary \ref{co2}. 
 
(e) Suppose that $S\le q-\beta A$, case (b) (Lemma \ref{l4}) takes place and $d<q-S<V_2$ (see Fig.\ref{fig4}). Then situation is similar to (b). Starting from any point $(y_0=b,~v_0\in[\beta A,A))$, at most $\varphi(\varphi(v_0))$ is such that the further trajectory never touches the axis $y=0$, the mappings $\varphi$ and $\tilde\varphi$ coincide on the domain $[q-S,A)$ and the required statements follow from Corollary \ref{co2}. 
\blacksquare 
 
Corollary \ref{cor_simple_cycle} is now obvious. 
 
\begin{conse}\label{co4} If a clipped $k$-cycle exists then an unclipped $k$-cycle exists, too. (See (\ref{ea1}).)\end{conse} 
 
\underline{Proof.} As is clear from the proof of 
Theorem~\ref{prop_stab}, $0<S\le q-\beta A$ and 
$\varphi(q-S)=\Phi^k(q-S)$. To put it differently, the domain of 
$\Phi^k$ is non-empty, so that the corresponding stationary point 
$V_1$ or $V_2$ (Fig.\ref{fig4}) does exist and defines the unclipped 
$k$-cycle. \blacksquare 
 
\begin{conse}\label{co5} The order of a clipped cycle can be $N$ or $N+1$ only (see (\ref{ee1})). \end{conse} 
 
\underline{Proof.} Suppose all parameters are fixed, apart from $b$. For very small values of $b$, obviously, only a clipped $N$-cycle is realised. Conditions when a clipped $(N+1)$-cycle exists, are left till the next subsecion. 
 
Suppose $N>1$. When we increase $b$, $k$-cycles with $k<N$ appear: see Fig.\ref{fig3}. If $b$ is close to $\frac{\beta^k}{1-\beta^k}-q$ then the $k$-cycle has a very short continuous part. From the proof of Lemma \ref{l5}, we have 
  $$\lim_{b\to\frac{\beta^k}{1-\beta^k}-q} s_0=0 \mbox{ and } \lim_{b\to\frac{\beta^k}{1-\beta^k}-q} y(s_0)=\frac{\beta^k}{1-\beta^k}.$$ 
(See (\ref{ea16},\ref{ea17}). Therefore, using Lemma \ref{l5}(a) we conclude that all $k$-cycles remain unclipped indeed. See also Corollary \ref{co3}. \blacksquare 
 
{\bf Effects of the router buffer $b$.} 
 
The goal of this subsection is to justify all the statements of 
Section~\ref{sec:cycles}. 
 
\underline{Case $A^*_{N+1}<q$} is trivial: see Fig.\ref{fig3}, Lemmas \ref{l5},\ref{l6}, Corollary \ref{co5} and its proof. 
 
\underline{Case $q\le q^*_{N+1}$.} According to Lemma \ref{l6}, here the $(N+1)$-cycle appears and becomes critical before it extincts at $b=A^*_{N+1}-q$. 
 
Consider the continuous trajectory (\ref{ea2}) staring from $(y_0=0,v_0=q)$: 
  $$\left\{\begin{array}{l} 
y(r)=e^{-r}+r-1; \\ v(r)=q+r. 
  \end{array}\right.$$ 
Clearly, there is $1-1$ correspondance between parameters $r$ and $b$ given by equation 
  \begin{equation}\label{ea22} 
e^{-r}+r-1=b. 
  \end{equation} 
The $(N+1)$-cycle cannot be realised if 
  $$\beta^N(q+r+1)<y(r)+q=e^{-r}+r-1+q=b+q.$$ 
Let us study the difference 
  \begin{equation}\label{ea23} 
\Delta(r)\defi e^{-r}+r-1+q-\beta^N(q+r+1). 
  \end{equation} 
Since $\frac{d\Delta(r)}{dr}=1-e^{-r}-\beta^N$, this difference has a minimum at $r=-\ln(1-\beta^N)$ (corresponding to $b=C$, see (\ref{ee3}) ) which equals 
  $$q(1-\beta^N)-2\beta^N-(1-\beta^N)\ln(1-\beta^N)=(1-\beta^N)(q-D),$$ 
see (\ref{ee2}). Since the critical $(N+1)$-cycle exists, we are sure that $q\le D$ and the values $\underline{b}$ and $\bar b$ (\ref{ee19}) are well defined. These equal the minimal and the maximal values providing $\Delta(r(b))=0$. Here and below, $r(b)$ is the positive solution to (\ref{ea22}). Note that the clipped $(N+1)$-cycle appears when $b=\underline{b}$ and becomes critical at $b=b_{0,N+1}$. The value $\bar b$ does not play any role because $\bar b\ge b_{0,N+1}$. 
 
\underline{Case $q^*_{N+1}<q\le A^*_{N+1}$.} Here the $(N+1)$-cycle cannot be critical (Lemma \ref{l6}). According to Lemma \ref{l5}, it also cannot be unclipped because unclipped cycle becomes critical when $b$ decreases. Sometimes $(N+1)$-cycles are not realised at all. Firstly, the latter happens if $D<q$. But even if $D\ge q$, it can happen that $\underline{b}>A^*_{N+1}-q$, so that the $(N+1)$-cycle does not exist in view of Corollary \ref{co4}. 
 
\begin{lemma}\label{l7} Suppose $q^*_{N+1}<q\le A^*_{N+1}$. 
 
(a) For a given value of $b$, the clipped $(N+1)$-cycle exists iff $\Delta(r(b))\le 0$ and $b\le A^*_{N+1}-q$. 
 
(b) $\Delta(r(A^*_{N+1}-q))>0$. 
 
(c) Suppose that $D\ge q$. Then $\underline{b}>A^*_{N+1}-q$ iff $C>A^*_{N+1}-q$; $\bar b<A^*_{N+1}-q$ iff $C<A^*_{N+1}-q$.\end{lemma} 
 
\underline{Proof.} (a) The necessity is obvious: see Corollary \ref{co4} and Fig.\ref{fig3}. 
 
Suppose $\Delta(r(b))\le 0$ and $b\le A^*_{N+1}-q$. For the unclipped $(N+1)$-cycle, the minimal value of $y$ is negative; let us denote the corresponding minimal value of $v$ by $\hat v$. Then, starting from $(y_0=0,v_0=q)$, the trajectory (\ref{ea2}) reaches the level $y=b$, and, after $(N+1)$ instant reductions of $v$, reaches point $(y=b,v<\hat v)$. After that, the trajectory goes down up to the axis $y=0$, and the clipped $(N+1)$-cycle is well defined. 
 
(b) Value $b=A^*_{N+1}-q$ is the largest buffer size when the unclipped $(N+1)$-cycle exists: see Fig.\ref{fig3}. The corresponding minimal value $y_{min}$ is negative and, starting from $(y_0=y_{min},v_0=y_{min}+q)$ trajectory (\ref{ea2}) reaches level $y=b$ at such value of $v$ that $\beta^Nv=b+q$. Therefore, starting from $(y_0=0,v_0=q)$, trajectory (\ref{ea2}) reaches level $y=b$ at a smaller value of $v$, and smaller than $(N+1)$ reductions of $v$ are needed, meaning that $\Delta(r(b))>0$. 
 
(c) Obviously, $\underline{b}\le C\le\bar b$. Thus the necessity is trivial. The sufficiency follows from (b) because $A^*_{N+1}-q\notin[\underline{b},\bar b]$. \blacksquare 
 
\begin{conse}\label{co6} In case $q^*_{N+1}<q\le A^*_{N+1}$, $q\le D$, the value of $C$ cannot equal $A^*_{N+1}-q$. \end{conse} 
 
The proof follows directly from statement (b), Lemma \ref{l7}. 
 
\begin{conse}\label{co7} Suppose $N$ is fixed. 
 
(a) For all $q\in(q^*_{N+1},A^*_{N+1}]$ the value of $N$ remains unchanged. 
 
(b) If $D\ge A^*_{N+1}$ then $\forall q\in(q^*_{N+1},A^*_{N+1}]$ $C>A^*_{N+1}-q$. 
 
(c) If $q^*_{N+1}<D<A^*_{N+1}$ then either $C>A^*_{N+1}-D$ and $\forall q\in (q^*_{N+1},D]$ $C>A^*_{N+1}-q$, or $C<A^*_{N+1}-D$ and $\forall q\in (q^*_{N+1},D]$ $C<A^*_{N+1}-q$. 
 
(d) If $\beta\in(0,1)$ varies, equality $C=A^*_{N+1}-D$ can hold only in the area where $D\le q^*_{N+1}$. 
\end{conse} 
 
\underline{Proof.} (a) The assertion follows from inequalities 
  $$\frac{\beta^N}{1-\beta^N}\le q^*_{N+1}<A^*_{N+1}\le\frac{\beta^{N-1}}{1-\beta^{N-1}},$$ 
see Lemma \ref{l2}(d) and Lemma \ref{l6}(b). As usual, $\frac{\beta^{N-1}}{1-\beta^{N-1}}=+\infty$ if $N=1$. 
 
(b) Clearly, if $q=A^*_{N+1}=\min\{A^*_{N+1},D\}$ then $C>0=A^*_{N+1}-q$. If we decrease $q$ up to $q^*_{N+1}$, the values of $C$ and $A^*_{N+1}$ remain unchanged and situation $C=A^*_{N+1}-q$ is excluded due to Corollary \ref{co6}. 
 
(c) The proof is similar to (b): take $q=D=\min\{A^*_{N+1},D\}$ and reduce its value. 
 
(d) In case $C=A^*_{N+1}-D$ and $D>q^*_{N+1}$ we have a contradiction to (c). \blacksquare 
 
One can show that different situations studied in Lemma \ref{l7} and Corollaries \ref{co6} and \ref{co7} can really take place. 
 
%
%
%
 
Theorem~\ref{pro2} follows directly from Section~\ref{sec:cycles}. 
 
\underline{Proof of Proposition \ref{pro3}.} According to definition (\ref{ee7}), $\delta=\frac{\beta(1+2\beta-\tau)}{1-\beta^2}$, where $\tau$ solves equation 
  $$\frac{\tau(1+\beta)}{1+2\beta+\beta\tau}=1-e^{-\tau}.$$ 
The both functions on the left and on the right increase from zero, and $\tau$ is smaller than $\theta$ which solves equation $\frac{\theta(1+\beta)}{1+2\beta+\beta\theta}=1$, ie $\tau<\theta=2\beta+1$. Now 
  $$\frac{\tau(1+\beta)}{1+2\beta+\beta\tau}<1-e^{-(2\beta+1)}\Longrightarrow \tau<\frac{(1+2\beta)(1-e^{-(2\beta+1)})}{1+\beta e^{-(2\beta+1)}}$$ 
and 
  $$\delta>\frac{\beta}{1-\beta^2}\cdot\frac{(1+2\beta)(\beta+1)e^{-(2\beta+1)}}{1+\beta e^{-(2\beta+1)}}\to\infty \mbox{ as } \beta\to 1.$$ 
\blacksquare 
 
\underline{Proof of Theorem~\ref{th_pareto}.} 
 
First we consider the case $b \in [0,b_{0,1}]$. In this case, the 
cycle is clipped or critical (see Figure~\ref{fig:ClipCycle}). 
According to Condition (b) of Theorem~\ref{pro2}, if $q>A^*_2$ the 
cycle does not have multiple jumps for any size of the buffer. 
Without loss of generality, we assume that the zero time moment 
corresponds to the time moment just after the jump (Point A). Recall 
that we denote the transformed time by $s$ and the original time by 
$t$. We denote by $S_A$ the transformed time when the system reaches 
point A, by $S_B$ the transformed time when the system reaches point 
B, and so on. Without loss of generality, we assume that $S_A=0$. We 
also use the notation: $S_{AB}=S_B-S_A=S_B$, $S_{BC}=S_C-S_B$, and 
so on. 
 
From (\ref{ea2}) we have 
$$ 
y(S_C+u)=y_{CD}(u)=e^{-u}+(u-1), \quad \mbox{for} \quad u \in 
[0,S_{CD}], 
$$ 
so that 
$$ 
y(S_D)=e^{-S_{CD}}+S_{CD}-1=b. 
$$ 
 
We note that $v(S_C)=q$. Consequently, $v(S_D)=q+S_{CD}$, 
$v(S_E)=q+S_{CD}+1$ and $v(S_A)=\beta(q+S_{CD}+1)$. Again, from 
(\ref{ea2}) we have 
$$ 
y(s)=(1+q+y(S_A)-v(S_A))e^{-s}+s-1+v(S_A)-q, 
$$ 
and 
$$ 
y(S_B)=[y(S_A)+1+q-v(S_A)]e^{-S_{AB}}+[S_{AB}-1] +v(S_A)-q=0. 
$$ 
Thus, we have the following equation for $S_{AB}$ 
$$ 
[b-\beta S_{CD}+(1-\beta)(1+q)]e^{-S_{AB}}+S_{AB} 
$$ 
$$ 
+\beta S_{CD} -(1-\beta)(1+q) = 0. 
$$ 
 
Now, we can calculate the cycle duration in the original and 
transformed times. Denote these quantities by $T_{cycle}$ and 
$S_{cycle}$, respectively. Note that $S_{cycle}=s_1+1$ (see 
(\ref{ea6}) with $k=1$). From equation $v(S_E)=v(S_A)+S_{cycle}$ 
we obtain 
$$ 
S_{cycle}=(1-\beta)(q+S_{CD}+1), 
$$ 
and, consequently, 
$$ 
T_{cycle}=\int_0^{T_{cycle}}dt 
=\int_0^{S_{cycle}}\left(T+\frac{x(s)}{\mu}\right)ds 
=TS_{cycle}+\frac{m}{\mu} \left(\frac{B}{m}+\int_{S_A}^{S_B}y(s)ds 
+\int_{S_C}^{S_D}y(s)ds\right). 
$$ 
Next, we calculate the average queue size 
$$ 
\bar{x}=\frac{1}{T_{cycle}}\int_0^{T_{cycle}}x(t)dt 
=\frac{1}{T_{cycle}}\int_0^{S_{cycle}}x(s)\left(T+\frac{x(s)}{\mu}\right)ds 
$$ 
$$ 
=\frac{1}{T_{cycle}}\left[ 
mT\left(\int_{S_A}^{S_B}y(s)ds+\int_{S_C}^{S_D}y(s)ds\right)\right. 
\left.+\frac{m^2}{\mu}\left(\int_{S_A}^{S_B}y^2(s)ds 
+\int_{S_C}^{S_D}y^2(s)ds\right)+B\left(T+\frac{B}{\mu}\right)\right]. 
$$ 
Now we calculate the average sending rate 
$$ 
\bar{\lambda}=\frac{1}{T_{cycle}}\int_0^{T_{cycle}}\lambda(t)dt 
$$ 
Using (\ref{lambda}), we have 
$$ 
\bar{\lambda}=\frac{1}{T_{cycle}}\int_0^{T_{cycle}} 
\frac{w(t)}{T+x(t)/\mu}dt 
=\frac{m}{T_{cycle}}\int_0^{S_{cycle}}v(s)ds 
$$ 
$$ 
=\frac{m}{T_{cycle}}\int_0^{S_{cycle}} 
\left(\beta(q+1+S_{CD})+s\right)ds =\frac{m}{T_{cycle}} \frac{1}{2} 
(1-\beta^2)(q+1+S_{CD})^2. 
$$ 
For the calculation of the average goodput we use the following 
formula: 
$$ 
\bar{g}=\frac{1}{T_{cycle}}\left[\int_{T_A}^{T_D}\lambda(t)dt 
+\mu\left(T+\frac{B}{\mu}\right)\right] 
=\frac{m}{T_{cycle}}\left[\int_{S_A}^{S_D}v(s)ds+q+b\right]. 
$$ 
In case $b \in (b_{0,1},\infty)$ the cycle is unclipped. 
Consequently, the calculations of the average quantities are more 
straightforward than in the previous case and are based on the 
knowledge of only one parameter $S_{cycle}$. \blacksquare 
 
\noindent \underline{Proof of Proposition~\ref{pro5}.} 
 
If $B \to \infty$ (equivalently, $b \to \infty$), then $s_1 \to 
\infty$ (see equation (\ref{ea6})). According to 
Theorem~\ref{th_pareto}, we have 
$$ 
T_{cycle}=\frac{m}{\mu}\left[q(s_1+1)+\int_0^{s_1}y(s)ds+b\right] 
$$ 
$$ 
=\frac{m}{\mu}\left[1+2b+2q-s_1-(1+b+q-\frac{\beta(s_1+1)}{1-\beta})e^{-s_1} 
+\frac{s_1^2}{2}+\frac{\beta(s_1^2-1)}{1-\beta}\right], 
$$ 
and, consequently, 
$$ 
\Delta=\mu\frac{\frac{1+\beta}{2(1-\beta)}(2s_1+1)-1-2b-2q+s_1+(1+b+q-\frac{\beta(s_1+1)}{1-\beta})e^{-s_1}+\frac{\beta}{1-\beta} 
}{1+2b+2q-s_1-(1+b+q-\frac{\beta(s_1+1)}{1-\beta})e^{-s_1}-\frac{\beta}{1-\beta}+\frac{1+\beta}{2(1-\beta)}s_1^2} 
$$ 
$$ 
\sim 
\frac{(2+\frac{2(1-\beta)}{1+\beta})s_1}{s_1^2}=\frac{4}{1+\beta}\frac{1}{s_1}\to 
0+, \quad \mbox{as} \quad s_1\to \infty. 
$$ 
\blacksquare 
 
\noindent \underline{Proof of Proposition~\ref{pro4}.}  (a) Suppose 
$N$ is fixed and $q=\frac{\mu T}{m}$ changes, i.e., increases 
starting from $\frac{\beta^N}{1-\beta^N}$. Using 
(\ref{ee4}),(\ref{ee5})  and omitting for brevity $N$ as the power 
and the index, we obtain: 
  $$\frac{dB_0}{dm}=m\frac{db_0}{d\theta}\cdot \frac{d\theta}{dq}\cdot\frac{dq}{dm}+b_0$$ 
  $$=m\left[\frac{1-e^{-\theta}-\theta e^{-\theta}}{(1-e^{-\theta})^2}\times \frac{\theta-1+e^{-\theta}}{\theta}\right] 
\left[\frac{1}{\frac{1-e^{-\theta}-\theta 
e^{-\theta}}{\theta(1-e^{-\theta})}+\frac{\beta}{1-\beta}}\right]\left[-\frac{\mu 
T}{m^2}\right]$$ 
 $$+\frac{\theta}{1-e^{-\theta}}-\ln\frac{\theta}{1-e^{-\theta}}-1.$$ 
We used the implicit differentiation theorem for 
$\frac{d\theta}{dq}$. Note that $\frac{\mu T}{m}=q$ and express $q$ 
using (\ref{ee4}): 
  $$\frac{dB_0}{dm}=\left[\frac{\theta}{1-e^{-\theta}}-\ln\frac{\theta}{1-e^{-\theta}}-1\right]$$ 
  $$-\left[ \frac{(1-e^{-\theta}-\theta e^{-\theta})(\theta-1+e^{-\theta})}{\theta(1-e^{-\theta})^2}\times 
\frac{\ln\frac{\theta}{1-e^{-\theta}}+\frac{\beta(\theta+1)}{1-\beta}}{\frac{1-e^{-\theta}-\theta 
e^{-\theta}}{\theta(1-e^{-\theta})}+\frac{\beta}{1-\beta}}\right].$$ 
The second square bracket, $f(\frac{\beta}{1-\beta})$, is a 
monotonous function of $\frac{\beta}{1-\beta}$. 
 
($\alpha$) If $(\theta+1)\frac{1-e^{-\theta}-\theta 
e^{-\theta}}{\theta(1-e^{-\theta})}-\ln\frac{\theta}{1-e^{-\theta}}\ge 
0$ then $f(\cdot)$ does not decrease and hence 
  $$\frac{dB_0}{dm}\le \frac{\theta}{1-e^{-\theta}}-\ln\frac{\theta}{1-e^{-\theta}}-1-f(0)$$ 
  $$=\frac{\theta}{1-e^{-\theta}}-\ln\frac{\theta}{1-e^{-\theta}}-1-\frac{\ln\frac{\theta}{1-e^{-\theta}}\cdot(\theta-1+e^{-\theta})}{(1-e^{-\theta})}$$ 
$$=\frac{\theta}{1-e^{-\theta}}-1-\frac{\theta}{1-e^{-\theta}}\ln\frac{\theta}{1-e^{-\theta}}=\gamma-1-\gamma\ln\gamma<0,$$ 
because $\gamma\defi\frac{\theta}{1-e^{-\theta}}\in(1,\infty)$ for 
$\theta>0$ and function $\gamma-1-\gamma\ln\gamma$ has the maximum 
which is equal to zero at $\gamma=1$. 
 
($\beta$) If 
  \begin{equation}\label{estar} 
(\theta+1)\frac{1-e^{-\theta}-\theta 
e^{-\theta}}{\theta(1-e^{-\theta})}-\ln\frac{\theta}{1-e^{-\theta}}<0 
  \end{equation} 
then $f(\cdot)$ decreases and hence 
  $$\frac{dB_0}{dm}<\frac{\theta}{1-e^{-\theta}}-\ln\frac{\theta}{1-e^{-\theta}}-1-\lim_{y\to\infty} f(y)$$ 
  $$=\frac{\theta}{1-e^{-\theta}}-\ln\frac{\theta}{1-e^{-\theta}}-1-\frac{(1-e^{-\theta}-\theta e^{-\theta})(\theta-1+e^{-\theta})(\theta+1)}{\theta(1-e^{-\theta})^2}.$$ 
Using (\ref{estar}), we have 
  $$\frac{dB_0}{dm}<\frac{\theta}{1-e^{-\theta}}-\frac{(\theta+1)(1-e^{-\theta}-\theta e^{-\theta})}{\theta(1-e^{-\theta})}-1$$ 
  $$-\frac{(1-e^{-\theta}-\theta e^{-\theta})(\theta-1+e^{-\theta})(\theta+1)}{\theta(1-e^{-\theta})^2}$$ 
  $$=\frac{3e^{-\theta}+\theta^2 e^{-\theta}+\theta e^{-\theta}-2-e^{-2\theta}}{(1-e^{-\theta})^2}<0, \mbox{ if } \theta>0.$$ 
Indeed, consider function $g(\theta)=3e^{-\theta}+\theta^2 
e^{-\theta}+\theta e^{-\theta}-2-e^{-2\theta}$. Clearly $g(0)=0$; 
  $$\frac{dg}{d\theta}=e^{-\theta}[\theta+2 e^{-\theta}-2-\theta^2];$$ 
  $$\left.\frac{dg}{d\theta}\right|_{\theta=0}=0;~~~~~\frac{d[\theta+2e^{-\theta}-2-\theta^2]}{d\theta}=1-2e^{-\theta}-2\theta<0$$ 
because the latter function decreases starting from $-1$ at 
$\theta=0$. 
 
Note that 
  \begin{equation}\label{e2star} 
\frac{dB_0}{dm}\to 0- \mbox{ as } \theta\to 0+. 
  \end{equation} 
 
(b) Obviously, without loss of generality we can put $N=1$ and prove 
that $B_{0,N}$ increases as $\beta\in(0,1)$ decreases. Like 
previously, we omit $N$ as the power and the index. Now again using 
the implicit differentiation theorem we obtain 
  $$\frac{dB_0}{d\beta}=m\frac{db_0}{d\theta}\cdot \frac{d\theta}{d\beta}$$ 
  $$=m\left[\frac{(1-e^{-\theta}-\theta e^{-\theta})(\theta-1+e^{-\theta})}{(1-e^{-\theta})^2\theta}\right]\left[\frac{-\frac{1+\theta}{(1-\beta)^2}}{\frac{1-e^{-\theta}-\theta e^{-\theta}}{\theta(1- e^{-\theta})}+\frac{\beta}{1-\beta}}\right]<0.$$ 
\blacksquare 
 
\noindent \underline{Proof of Corollary~\ref{co8}.} The first part 
follows directly from Proposition~\ref{pro4}, if we notice that $N$ 
remains unchanged on intervals $m\in\left[\frac{\mu 
T(1-\beta^{N-1})}{\beta^{N-1}},\frac{\mu 
T(1-\beta^N)}{\beta^N}\right)$ and increases by $1$ at points 
$m_{i+1}$. 
 
When $m\to m_N-0$, $q$ approaches $\frac{\beta^N}{1-\beta^N}$ and 
$\theta_N$ goes to zero (see (15)). According to (\ref{ee5}), 
$b_{0,N}\to 0+$, hence $B_{0,N}=mb_{0,N}\to 0+$. Equality 
(\ref{e2star}) implies that $\frac{dB_{0,N}}{dm}\to 0-$. 
 
Suppose $m\to 0+$, $q\to\infty$, $N=1$, $\theta_1\to\infty$. Then 
$\frac{\ln\frac{\theta_1}{1-e^{-\theta_1}}}{\theta_1}\to 0$ and 
$\frac{q}{\theta_1}\to\frac{\beta}{1-\beta}$ according to 
(\ref{ee4}). Therefore 
  $$B_{0,1}=mb_{0,N}=\frac{\mu T}{q}\theta_1\left(\frac{b_{0,1}}{\theta_1}\right)\to\frac{\mu T(1-\beta)}{\beta}.$$ 
 
Consider $B_{0,N}(m_{N-1})$, ie put 
$q=\frac{\beta^{N-1}}{1-\beta^{N-1}}$ and study equations 
(\ref{ee4}),(\ref{ee5}). Let $\theta_N$ be the positive solution to 
\begin{equation}\label{e3star} 
\ln\frac{\theta}{1-e^{-\theta}}+\frac{\beta^N}{1-\beta^N}\cdot\theta=\frac{\beta^{N-1}}{1-\beta^{N-1}}-\frac{\beta^N}{1-\beta^N}; 
\end{equation} 
then 
  $$B_{0,N}(m_{N-1})=\frac{\mu T(1-\beta^{N-1})}{\beta^{N-1}}\left[\frac{\theta_N}{1-e^{-\theta_N}}-\ln\frac{\theta_N}{1-e^{-\theta_N}}-1\right].$$ 
Obviously, $\lim_{N\to\infty}\theta_N=0$, hence, directly from 
(\ref{e3star}) we obtain: 
  $$\lim_{N\to\infty}\left[\frac{\ln\frac{\theta_N}{1-e^{-\theta_N}}}{\theta_N}\cdot\frac{\theta_N}{\beta^N}+\frac{\theta_N}{1-\beta^N}\right] 
  =\frac{1}{2}\lim_{N\to\infty}\frac{\theta_N}{\beta^N}=\lim_{N\to\infty}\left[\frac{\beta^{-1}}{1-\beta^{N-1}}-\frac{1}{1-\beta^N}\right]=\frac{1-\beta}{\beta}$$ 
and finally 
  $$\lim_{N\to\infty} m_{N-1}B_{0,N}(m_{N-1})=\lim_{N\to\infty}\left[\left.\left(\frac{\theta_N}{1-e^{-\theta_N}}-\ln\frac{\theta_N}{1-e^{-\theta_N}}-1\right)\right/ \theta_N^2\right]$$ 
  $$\times\lim_{N\to\infty}\left(\frac{\theta_N}{\beta^N}\right)^2\beta^2(\mu T)^2\lim_{N\to\infty}(1-\beta^{N-1})^2 
=\frac{1}{8}\left[\frac{2(1-\beta)}{\beta}\right]^2\beta^2(\mu T)^2=\frac{1}{2}(1-\beta)^2(\mu T)^2.$$ 
\blacksquare

\end{document}